\documentclass[aps,pra,reprint,onecolumn,superscriptaddress,floatfix]{revtex4-1}

\usepackage{natbib}
\usepackage{anysize}
\usepackage{graphicx}
\usepackage{times}
\usepackage{amsmath,amssymb}
\usepackage{dcolumn}% Align table columns on decimal point
\usepackage{bm}% bold math
\usepackage{color}
\usepackage[colorlinks={true},dvipdfm]{hyperref}  % use with DVI
\hypersetup{citecolor={blue}, filecolor={blue}, linkcolor={blue}, urlcolor={blue}}

\begin{document}

\title{Exploring the trade-off between fidelity- and time-optimal control of quantum unitary transformations}
\author{Katharine W. Moore Tibbetts}
\affiliation{Department of Chemistry, Princeton University, Princeton, NJ 08544, USA}
\affiliation{Department of Scalable \& Secure Systems Research, Sandia National Laboratories, Livermore, CA 94550, USA}
\author{Constantin Brif}
\author{Matthew D. Grace} 
\affiliation{Department of Scalable \& Secure Systems Research, Sandia National Laboratories, Livermore, CA 94550, USA}
\author{Ashley Donovan}
\author{David L. Hocker}
\author{Tak-San Ho}
\affiliation{Department of Chemistry, Princeton University, Princeton, NJ 08544, USA}
\author{Re-Bing Wu}
\affiliation{Department of Automation and Center for Quantum Information Science and Technology, Tsinghua University, Beijing, 100084, China}
\author{Herschel Rabitz}
\affiliation{Department of Chemistry, Princeton University, Princeton, NJ 08544, USA}

%\date{\today}

\begin{abstract}
Generating a unitary transformation in the shortest possible time is of practical importance to quantum information processing because it helps to reduce decoherence effects and improve robustness to additive control field noise. Many analytical and numerical studies have identified the minimum time necessary to implement a variety of quantum gates on coupled-spin qubit systems. This work focuses on exploring the Pareto front that quantifies the trade-off between the competitive objectives of maximizing the gate fidelity $\mathcal{F}$ and minimizing the control time $T$. In order to identify the critical time $T^{\ast}$, below which the target transformation is not reachable, as well as to determine the associated Pareto front, we introduce a numerical method of Pareto front tracking (PFT). We consider closed two- and multi-qubit systems with constant inter-qubit coupling strengths and each individual qubit controlled by a separate time-dependent external field. Our analysis demonstrates that unit fidelity (to a desired numerical accuracy) can be achieved at any $T \geq T^{\ast}$ in most cases. However, the optimization search effort rises superexponentially as $T$ decreases and approaches $T^{\ast}$. Furthermore, a small decrease in control time incurs a significant penalty in fidelity for $T < T^{\ast}$, indicating that it is generally undesirable to operate below the critical time. We investigate the dependence of the critical time $T^{\ast}$ on the coupling strength between qubits and the target gate transformation. Practical consequences of these findings for laboratory implementation of quantum gates are discussed.
\end{abstract}
\maketitle

\section{Introduction}
\label{sec:intro}

The goal of controlling the dynamics of a quantum system in order to generate a target unitary transformation is both of fundamental interest and directly applicable to implementation of logic operations in quantum information processing \cite{NielsenChuang2000}. Two strategies are commonly employed to design control fields that enact the desired evolution: (i) geometric techniques for analytically constructing control pulse sequences \cite{Khaneja2001, *Khaneja2002, KhanejaHeitmann2007, BoscainChitour2005} and (ii) numerical methods employing optimal control theory (OCT) \cite{Brif2010NJP, Brif2012ACP, BalintKurti2008, DAlessandro2007}. The effectiveness of the OCT approach has been demonstrated both for the idealized case of closed quantum systems undergoing unitary evolution and for open quantum systems whose dynamics are affected by coupling to the environment \cite{Brif2010NJP, Brif2012ACP}. Control fields producing quantum unitary transformations with a high fidelity have been successfully identified using OCT for a variety of models involving coupled-spin systems \cite{KhanejaReiss2005, SchulteSporl2005, GraceBrif2007JPB, *GraceBrif2007JMO, GraceDominy2010NJP, Schirmer2009JMO, TsaiChenGoan2009PRA, WeninPotz2008PRA, *WeninPotz2008PRB, *RoloffWeninPotz2009JCE, *RoloffWeninPotz2009JCTN, BurgarthMaruyama2010PRA, SchulteSporl2011JPB}, molecular systems \cite{PalaoKosloff2002, *PalaoKosloff2003, TeschVivieRiedle2002, *VivieRiedleTroppmann2007, GraceBrif2006}, and other physical realizations \cite{SporlSchulte2007, MontangeroCalarcoFazio2007PRL, SafaeiMontangero2009, WeninRoloffPotz2009, Nebendahl2009, Poulsen2010PRA, GoerzCalarcoKoch2011JPB, KhaniMerkel2012PRA}.

An important physical parameter for quantum computation is the control time $T$ required to generate a target quantum gate. In general, decreasing the control duration helps to reduce the effect of decoherence resulting from the interaction of a quantum system with the environment \cite{NielsenChuang2000}. Also, the gate error due to additive white noise in the control fields grows linearly with $T$ \cite{Brif:robustness}, which means that shorter control times will enhance the robustness to this type of noise. Due to these considerations, the problem of identifying control fields that enact a target quantum gate to a specified fidelity in the minimum time, called time-optimal control (TOC) \cite{Khaneja2001}, is important for practical quantum information processing. The minimum time required to implement a multi-qubit gate is also related to the gate's complexity expressed as the number of one- and two-qubit gates necessary to construct the target unitary transformation \cite{Nielsen2006}. TOC was originally formulated as a geometric problem of identifying the geodesic between two elements of the unitary group $\mathrm{U}(N)$ and solved using Lie group methods and Pontryagin's minimum principle \cite{Khaneja2001, Khaneja2002}. This analytical technique has been applied to identify control pulse sequences and associated values of the minimum time $T^{\ast}$ for generating quantum gates in two- and three-qubit NMR and other coupled-spin systems \cite{Khaneja2001, *Khaneja2002, KhanejaHeitmann2007, BoscainChitour2005, ZhangWhaley2003, Masanes2002}. An alternative approach is to solve variational equations for the optimal time-dependent Hamiltonian $H(t)$ under the constraint of finite energy using the quantum brachistochrone method \cite{Carlini2007, Carlini2011}, which can provide improved $T^{\ast}$ values for some systems \cite{Carlini2011}. Other OCT-based algorithms utilize additional cost terms to penalize the control duration \cite{MishimaYamashita2009a, *MishimaYamashita2009b, Lapert2012PRA}. In addition, several numerical studies have identified $T^{\ast}$ values approaching or even improving upon analytical results when employing OCT to design optimal fields \cite{KhanejaReiss2005, SchulteSporl2005, KoikeOkudaira2010}.

In this work, we incorporate the goal of TOC by considering a more general problem of quantum Pareto optimization \cite{RajWu2008b} for the objectives of maximizing the gate fidelity $\mathcal{F}$ and minimizing the control time $T$. These two objectives compete with each other when $T < T^{\ast}$, with the relationship between the best simultaneously achievable values of $\mathcal{F}$ and $T$ constituting the Pareto front. Previously, Pareto optimization has been explored both theoretically and experimentally for the goal of discriminating between similar quantum systems \cite{BeltraniGhoshRabitz2009, RothGuyonRoslund2009, *RoslundRothGuyon2011}, as well as theoretically for maintaining persistent field-free control \cite{AnsonBeltraniRabitz2011}. For TOC, some numerical simulations have sought the best fidelity value attainable at a given $T$ \cite{SchulteSporl2005, KoikeOkudaira2010}, but without explicitly investigating the Pareto front, especially in regions of high fidelity that are important for quantum information processing. We explore these Pareto fronts with the goal of identifying the relationship between the simultaneously achievable $\mathcal{F}$ and $T$ values, as well as their dependence on the target unitary transformation and inter-qubit coupling strength. In order to numerically implement this analysis, we introduce the Pareto front tracking (PFT) algorithm, which sequentially (a) makes a small variation in one objective (here, decreases the control time $T$) and (b) searches for a control field that optimizes the second objective (here, maximizes the gate fidelity $\mathcal{F}$). Unlike the procedures described in Refs.~\cite{RajWu2008b, BeltraniGhoshRabitz2009}, the PFT method does not simultaneously optimize both control objectives, and thus may be less computationally expensive, especially when the value of one objective (here, $T$) is easily varied, but not easily optimized using OCT. The PFT algorithm is applicable to any such pair of objectives, for example, control field fluence and fidelity. In this work, we consider only the objectives of minimizing $T$ and maximizing $\mathcal{F}$.

In addition to exploring the region of the Pareto front corresponding to $T < T^{\ast}$ (where the maximum attainable fidelity is limited to values below $1$), it is also of interest to understand how the optimization search effort (quantified as the number of algorithmic iterations needed to reach the optimum to a desired numerical accuracy) is affected when approaching the critical value $T^{\ast}$ from $T > T^{\ast}$. For the objective of generating unitary transformations, the search effort was found to exhibit large variations with respect to the Hamiltonian structure \cite{MooreChakrabarti2011}. In this work, we observe that the search effort to find optimal control solutions grows very rapidly as $T$ decreases towards $T^{\ast}$, even though a fidelity value arbitrarily close to the maximum $\mathcal{F} = 1$ is, in principle, attainable at any $T \geq T^{\ast}$. To facilitate the understanding of this behavior, we consider properties of the optimal control landscape, which is defined by the functional dependence of the physical objective (here, the gate fidelity $\mathcal{F}$) on the applied controls \cite{Brif2010NJP, Brif2012ACP, RabitzHsiehRosenthal2004, RabitzHsiehRosenthal2005PRA, GirardeauKoch1998, ChakrabartiRabitz2007review}. For controllable quantum systems \cite{Ramakrishna1995} with unconstrained control resources, the set of regular critical points on the unitary-transformation control landscape contains no local optima \cite{HsiehRabitz2008PRA, ChakrabartiRabitz2007review, HoDominyRabitz2009PRA}. This property of the landscape topology is directly relevant to the optimization behavior, since local optima may act as ``traps'' for a gradient-based search. The lack of traps on the control landscape has been verified with carefully conducted numerical simulations that ensured that no significant constraints were placed on the control fields \cite{MooreChakrabarti2011}. Since the goal of TOC inherently involves limiting an important control resource (specifically, the control time $T$), it is possible that the favorable landscape topology may break down as the critical time $T^{\ast}$ is approached (fortunately, as we will show, this possibility does not materialize). Furthermore, it is of interest to explore how the local structure of the control landscape changes near $T^{\ast}$ and how this change is related to the rise of the search effort in this region. The PFT method introduced in this work is well-suited for exploring the landscape regions in the vicinity of the maximum fidelity while approaching $T^{\ast}$ from $T > T^{\ast}$. In order to quantify how the local landscape structure changes upon approaching $T^{\ast}$, we employ metrics similar to those developed in Refs.~\cite{MooreHsiehRabitz2008JCP, MooreRabitz2011, MooreChakrabarti2011} and demonstrate their correlation with the search effort.

The remainder of this paper is organized as follows. Section~\ref{sec:background} presents the background and motivation for the current study, including the formulation of the optimal control problem, the optimization algorithm, metrics on the control landscape, model physical systems, the relationship between robustness to additive white control noise and control time, and the method for tracking the Pareto front. In Sec.~\ref{sec:2q}, we explore the fidelity-time Pareto fronts and the effect of control-time reduction on the search effort for two-qubit gates. The study is extended to three- and four-qubit gates in Sec.~\ref{sec:multiq}. Finally, Sec.~\ref{sec:conclusion} presents concluding remarks.

\section{Background and motivation}
\label{sec:background}

\subsection{Formulation of the control objective}
\label{sec:formulation}

We consider an $N$-level closed quantum system whose evolution is governed by the time-dependent Schr{\"o}dinger equation (in units where $\hbar = 1$):
\begin{equation}
\label{eq:Schro}
i \frac{\partial U(t,0)}{\partial t} = H(\{\varepsilon_k(t)\})U(t,0),
\quad U(0,0) \equiv \openone, 
\end{equation}
where $H(\{\varepsilon_k(t)\})$ is the Hamiltonian, $\{\varepsilon_k(t)\}$ are time-dependent external control fields, $U(t,0)$ is the unitary propagator (time-evolution operator) from time $t = 0$ to $t$, and $\openone$ is the identity operator. We will use the shorthand notation $U(t) \equiv U(t,0)$ for simplicity, where applicable. The propagator at some final time $T$ is denoted as $U_T \equiv U(T)$ and is a functional of the control fields: $U_T = U_T(\{\varepsilon_k(t)\})$. We assume linear (dipole-type) coupling to the control fields: 
\begin{equation}
\label{eq:Ham}
H(\{\varepsilon_k(t)\}) = H_0 + \sum_k \varepsilon_k(t) H_{\mathrm{c}}^{(k)} ,
\end{equation}
where $H_0$ is the field-free Hamiltonian and $\{H_{\mathrm{c}}^{(k)}\}$ are the control-Hamiltonian operators. The quantum system is assumed to be evolution-operator controllable, which means that any desired $U_T \in \mathrm{U}(N)$ [or, $U_T \in \mathrm{SU}(N)$ for a traceless Hamiltonian] can be generated through the Schr{\"o}dinger evolution (\ref{eq:Schro}) by some choice of control fields $\{\varepsilon_k(t)\}$ at a sufficiently large time $T$ \cite{Ramakrishna1995, Brif2012ACP, DAlessandro2007}. The necessary and sufficient condition for evolution-operator controllability is that the operators $\{ H_0, H_{\mathrm{c}}^{(k)} \}$ generate the Lie algebra $\mathrm{u}(N)$ [$\mathrm{su}(N)$ for a traceless Hamiltonian] \cite{Ramakrishna1995}.

In circuit-model quantum computing, the goal is to generate specified unitary transformations that implement desired logic operations on a system of qubits. The corresponding control objective is to guide the system's final-time propagator $U_T$ to match a specified unitary transformation $W$. A convenient mathematical formulation of this objective is to minimize the distance between $U_T$ and $W$:
\begin{equation}
\label{Jdef}
\mathcal{D}(U_T) = \|W-U_T\|_{\mathrm{HS}}^2
= 2 N - 2\mathrm{Re}\mathrm{Tr}(W^{\dagger}U_T) ,
\end{equation}
where $\| X \|_{\mathrm{HS}}^2 \equiv \mathrm{Tr}(X^{\dag} X)$ is the squared Hilbert-Schmidt norm. The desired minimum $\mathcal{D} = 0$ is achieved when $U_T = W$, and the maximum $\mathcal{D} = 4N$ corresponds to $U_T = -W$. It is often convenient to use the normalized distance $\tilde{\mathcal{D}}$:
\begin{equation}
\label{nf}
\tilde{\mathcal{D}}(U_T) = \frac{1}{4 N} \mathcal{D} 
= \frac{1}{2} - \frac{1}{2 N} \mathrm{Re}\mathrm{Tr}(W^{\dag}U_T) ,
\end{equation}
so that $\tilde{\mathcal{D}}$ takes values in the interval $[0,1]$. A commonly employed gate fidelity $\mathcal{F}$ is related to the distance as 
\begin{equation}
\label{eq:fidelity}
\mathcal{F}(U_T) = 1 - \tilde{\mathcal{D}} = \frac{1}{2} + \frac{1}{2 N} \mathrm{Re}\mathrm{Tr}(W^{\dag}U_T) .
\end{equation}
The minimum distance ($\mathcal{D} = \tilde{\mathcal{D}} = 0$) at $U_T = W$ corresponds to the maximum fidelity $\mathcal{F} = 1$. The functional dependence of the objective on the control fields, i.e., $\mathcal{D} = \mathcal{D}(\{\varepsilon_k(t)\})$ or, equivalently, $\mathcal{F} = \mathcal{F}(\{\varepsilon_k(t)\})$ determines the optimal control landscape.

OCT is often formulated by applying the variational principle to an objective functional, such as $\mathcal{F}$ (or, equivalently, $\mathcal{D}$), along with Lagrange multipliers to ensure satisfaction of the Schr\"odinger equation (\ref{eq:Schro}) as well as to impose a constraint on the control field fluence \cite{Brif2012ACP, PalaoKosloff2002, *PalaoKosloff2003, MooreChakrabarti2011}. An alternative approach \cite{HoRabitz2006JPPA, HoDominyRabitz2009PRA, MooreRabitz2011} is to consider small responses in the propagator $U(t)$ due to changes in the control fields $\delta\varepsilon_k$, subject to the Schr\"odinger equation:
\begin{equation}
\label{du}
i \frac{\partial}{\partial t} \delta U(t,0) = H(t) \delta U(t,0) + \delta H(t)U(t,0) , 
\end{equation}
with the initial condition $\delta U(0,0) = 0$. Equation (\ref{du}) can be integrated \cite{HoRabitz2006JPPA, HoDominyRabitz2009PRA} to give
\begin{equation}
\label{intdu}
\delta U(t,0) = -i \int_0^t \! d t'\, U(t,t')\delta H(t')U(t',0) .
\end{equation}
For the Hamiltonian of Eq.~(\ref{eq:Ham}), the variation is given by $\delta H(t) = \sum_k H_{\mathrm{c}}^{(k)} \delta \varepsilon_k(t)$; by using this result in Eq.~(\ref{intdu}), one obtains \cite{HoDominyRabitz2009PRA} the variation of the propagator $U_T$ with respect to the control $\varepsilon_k(t)$:
\begin{equation}
\label{dude}
 \frac{\delta U_T}{\delta\varepsilon_k(t)} = -i U_T H_{\mathrm{c}}^{(k)}(t),
\end{equation}
where $H_{\mathrm{c}}^{(k)}(t) = U^{\dag}(t)H_{\mathrm{c}}^{(k)}U(t)$. Combining Eqs.~(\ref{Jdef}) and (\ref{dude}), we obtain the desired functional derivative of the distance $\mathcal{D}(U_T)$ with respect to $ \varepsilon_k(t)$:
\begin{align}
 \frac{\delta \mathcal{D}}{\delta \varepsilon_k(t)} &= -2 \mathrm{Re}\mathrm{Tr} \left[ W^{\dag} \frac{\delta U_T}{\delta\varepsilon_k(t)} \right] \nonumber \\
&= -2 \mathrm{Im}\mathrm{Tr} \left[ W^{\dag} U_T H_{\mathrm{c}}^{(k)}(t) \right] . 
\label{dfde}
\end{align}

The critical points of the control landscape (also referred to as extremal solutions) are control fields $\{\varepsilon_k(t)\}$ that satisfy
\begin{equation}
\label{eq:crit-dyn}
\frac{\delta \mathcal{D}}{\delta \varepsilon_k(t)} = 0 , \quad 
\forall k \ \ \text{and} \ \ \forall t \in [0,T] . 
\end{equation}
Quantum control landscape theory has shown \cite{Brif2012ACP, ChakrabartiRabitz2007review} that when (a) the system is controllable, (b) no significant constraints are placed on the control fields, and (c) the Jacobian in Eq.~(\ref{dude}) is full-rank, \emph{dynamic} critical points satisfying Eq.~(\ref{eq:crit-dyn}) occur only at \emph{kinematic} critical points that satisfy $\nabla \mathcal{D}(U_T) = 0$. The values of the distance $\mathcal{D}$ at the critical points are $\mathcal{D} = 0, 4, 8, \ldots , 4 N$ \cite{HsiehRabitz2008PRA, HoDominyRabitz2009PRA}. The optimality of a critical point can be determined by inspecting eigenvalues of the Hessian matrix $\mathsf{H}(t,t') = \delta^2 \mathcal{D}/\delta \varepsilon(t)\delta\varepsilon(t')$ \footnote{Here, $\varepsilon(t)$ denotes the concatenation of all control functions in the set $\{ \varepsilon_k(t) \}$, so that the Hessian matrix $\mathsf{H}(t,t')$ is the concatenation of all blocks $\mathsf{H}_{k j}(t,t') = \delta^2 \mathcal{D}/\delta \varepsilon_k(t)\delta\varepsilon_j(t')$.}. Thus, the values $\mathcal{D} = 0$ and $\mathcal{D} = 4 N$ correspond to the global minimum (the Hessian is positive semidefinite) and global maximum (the Hessian is negative semidefinite), respectively. Moreover, it has been shown through analysis \cite{HsiehRabitz2008PRA, HoDominyRabitz2009PRA} and numerical simulations \cite{MooreChakrabarti2011} that, under the conditions (a)--(c) above, all the intermediate critical points (i.e., $\mathcal{D} = 4, \ldots , 4 N - 4$) have a saddle-point topology (the Hessian has positive, negative, and zero eigenvalues), meaning that no local maxima or minima exist on the landscape. However, when control resources are severely constrained (e.g., by limiting the control time $T$ as considered here), the trap-free landscape topology is no longer guaranteed.

Values of the distances and fidelity of Eqs.~(\ref{Jdef}), (\ref{nf}), and (\ref{eq:fidelity}) depend on the global phase of the transformation $U_T$. Since this global phase is physically irrelevant for a given quantum gate, a phase-independent version of the distance (or fidelity) is often employed instead \cite{HoDominyRabitz2009PRA}. In particular, a normalized phase-independent distance can be defined as \footnote{The normalization in Eq.~(\ref{ind-def}) is chosen so that the $\mathcal{G}$ takes values in the interval $[0,1]$, analogous to $\tilde{\mathcal{D}}$.}
\begin{equation}
\label{ind-def}
\mathcal{G}(U_T) = \frac{1}{2 N} \min_{\phi} \mathcal{D}(e^{i \phi} U_T) .
\end{equation}
The minimization over the global phase $\phi$ in Eq.~(\ref{ind-def}) can be easily carried out to obtain:
\begin{equation}
\label{ind}
\mathcal{G}(U_T) = 1 - \frac{1}{N} \left| \mathrm{Tr}(W^{\dag} U_T) \right| .
\end{equation}
Correspondingly, the minimum value $\mathcal{G} = 0$ is attained when $U_T = e^{i \phi} W$ for any phase $\phi$. Note that for a traceless Hamiltonian, when both $W$ and $U_T$ must be in SU($N$), the phase $\phi$ can take only discrete values corresponding to solutions of the equation $e^{i N \phi} = 1$. The topology of the control landscape for the distance $\mathcal{G}$ of Eq.~(\ref{ind}) is very similar to that for the distance $\mathcal{D}$ of Eq.~(\ref{Jdef}), and an analytical formula for the functional derivative $\delta \mathcal{G}/\delta \varepsilon_k(t)$ can be obtained analogous to Eq.~(\ref{dfde}) \cite{HoDominyRabitz2009PRA}. Optimization of the phase-independent distance $\mathcal{G}$ will be considered in Sec.~\ref{sec:2q2}, where we study how searches with different initial control fields converge to optimal solutions corresponding to different values of the global phase. Optimization of the phase-dependent distance $\tilde{\mathcal{D}}$ is considered in the remainder of this paper (of course, minimizing the distance $\tilde{\mathcal{D}}$ is equivalent to maximizing the fidelity $\mathcal{F}$).

\subsection{Optimization procedure for control of unitary transformations}
\label{sec:dmorph}

A variety of deterministic first-order algorithms, including the Krotov method \cite{PalaoKosloff2002, *PalaoKosloff2003, KoikeOkudaira2010, SchirmerFouquieres2011NJP}, GRAPE algorithm \cite{KhanejaReiss2005, SchulteSporl2005}, and D-MORPH (diffeomorphic modulation under observable-response-preserving homotopy) \cite{DominyRabitz2008JPA, MooreChakrabarti2011} have been employed for optimization in control of quantum unitary transformations. A recent work \cite{MooreChakrabarti2011} demonstrated that these algorithms share a common fixed point topology and common bounds on their convergence rates. Also, methods for comparison and benchmarking of various quantum control algorithms were presented in Ref.~\cite{Machnes2011PRA}. Second-order algorithms such as the Broyden--Fletcher--Goldfarb--Shanno (BFGS) quasi-Newton method \cite{Machnes2011PRA, FouquieresSchirmer:traps} and the Newton--Raphson method \cite{deFouquieres2012PRL} were also utilized recently in quantum optimal control. In this work, we employ the D-MORPH method.

In D-MORPH, a variable $s$ (referred to as the algorithmic index) is introduced to label the progression of the optimization, rendering the control fields: $\{ \varepsilon_k(t) \} \rightarrow \{ \varepsilon_k(s,t) \}$. The objective value $\mathcal{D}$ depends on $s$ through its functional dependence on the set of control fields, i.e., $\mathcal{D}(s) = \mathcal{D}(\{\varepsilon_k(s,t)\})$. Thus, the change in the objective value corresponding to a differential change $d s$ is given by $d \mathcal{D} = \left( \partial \mathcal{D}/\partial s \right) d s$, where
\begin{equation}
\label{dJds}
\frac{\partial\mathcal{D}}{\partial s} = \int_0^T \! d t \sum_k \frac{\delta \mathcal{D}}{\delta\varepsilon_k(s,t)}\frac{\partial \varepsilon_k(s,t)}{\partial s} .
\end{equation} 
As the goal is to minimize $\mathcal{D}$, we require that $\partial\mathcal{D}/\partial s \leq 0$, which is guaranteed when each control $\varepsilon_k(s,t)$ satisfies the differential equation 
\begin{equation}
\label{ode}
\frac{\partial\varepsilon_k(s,t)}{\partial s} = -\frac{\delta \mathcal{D}}{\delta\varepsilon_k(s,t)} .
\end{equation}
The functional derivative on the right-hand side of Eq.~(\ref{ode}) can be evaluated using Eq.~(\ref{dfde}). In numerical simulations, we determine control fields at each iteration by solving Eq.~(\ref{ode}) using a fourth-order Runge-Kutta integrator with a variable step size incorporated into MATLAB (routine \texttt{ode45}) \cite{matlab}. The initial set of fields $\{\varepsilon_k(0,t)\}$ is selected randomly for each optimization run, as described in Sec.~\ref{sec:syst} below. In the simulations, we evaluate the normalized distance $\tilde{\mathcal{D}}$ and specify the convergence threshold for the optimization. Specifically, the D-MORPH procedure described above is performed until either (a) the desired convergence criterion $\tilde{\mathcal{D}} \leq 10^{-8}$ is reached or (b) the improvement of the distance value satisfies $|\tilde{\mathcal{D}}(s + d s) - \tilde{\mathcal{D}}(s)| \leq 10^{-6} \tilde{\mathcal{D}}(s)$, indicating that an extremal value of $\tilde{\mathcal{D}}$ has been reached. Sufficient algorithmic iterations are allowed to prevent premature termination before reaching an extremal value of $\tilde{\mathcal{D}}$.

\subsection{Metrics of landscape structure}
\label{sec:metrics}

The issue of the search effort growth as $T \rightarrow T^{\ast}$ is important because the computational cost of identifying optimal control fields near or at the critical time may become prohibitively high, particularly as the number of controlled qubits increases. We are also interested in exploring the relationship between the number of D-MORPH algorithmic iterations needed to converge to an optimal solution and metrics that quantify local properties of the control landscape. One such metric is the \emph{path length} of the search trajectory for a D-MORPH optimization. The search starts out from an initial set of control fields $\{\varepsilon_k(0,t)\}$ at the algorithmic index value $s = 0$ and progresses in steps $s \rightarrow s + d s$ until the trajectory ends at a set of optimal control fields $\{\varepsilon_k^{\star}\} = \{\varepsilon_k(s^{\star},t)\}$ with $s = s^{\star}$. The path length $\Lambda(s)$ of the search trajectory from $s = 0$ to $s$ is defined as
\begin{equation}
\label{path}
\Lambda(s) = \int_0^s \! d s' \left\{ \frac{1}{T} \sum_k \int_0^T \! d t \left[\frac{\partial \varepsilon_k(s',t)}{\partial s'}\right]^2 \right\}^{1/2} . 
\end{equation}
The total path length traveled along the search trajectory to reach the optimum is $\Lambda^{\star} = \Lambda(s^{\star})$. In the absence of local traps, a D-MORPH search monotonically converges towards the optimum, resulting in a one-to-one mapping between the algorithmic index value $s$ and the objective value $\tilde{\mathcal{D}}$ for a given search trajectory. Therefore, it is possible to cast the path length as a function of the distance: $\Lambda = \Lambda(\tilde{\mathcal{D}})$. The value of $\Lambda(\tilde{\mathcal{D}})$ quantifies how much the control fields have to change from the initial guess $\{\varepsilon_k(0,t)\}$ to achieve the objective value $\tilde{\mathcal{D}}$.

In Ref.~\cite{MooreChakrabarti2011}, it was found that metrics of the local landscape structure near the optimum can be useful for predicting the search effort. In particular, one can employ the \emph{slope metric} $\Sigma (s)$, which is given by the $L_2$-norm of the landscape gradient $\nabla_{\varepsilon} \mathcal{D}$ evaluated at an algorithmic index value $s$:
\begin{equation}
\label{steep}
\Sigma(s) = \| \nabla_{\varepsilon} \mathcal{D}(s) \|_2 = \left\{ \sum_k \int_0^T \!\! d t \left[ \frac{\delta \mathcal{D}}{\delta\varepsilon_k(s,t)} \right]^2 \right\}^{1/2} \! .
\end{equation}
During the D-MORPH optimization, the derivative of the $k$th control field with respect to $s$ and the landscape gradient along this field are related by Eq.~(\ref{ode}). Substituting Eq.~(\ref{ode}) into Eq.~(\ref{path}) and using Eq.~(\ref{steep}), we obtain \footnote{Since the numerical implementation of the D-MORPH optimization employs a variable step size $d s$, this variation should be taken into account when evaluating the integral and differential in Eq.~(\ref{path-1})}:
\begin{equation}
\label{path-1}
\Lambda(s) = \frac{1}{\sqrt{T}} \int_0^s \! d s' \Sigma(s') \ \ \Leftrightarrow \ \ 
\Sigma(s) = \sqrt{T} \frac{d \Lambda(s)}{d s} .
\end{equation}
This simple relationship shows that the slope metric $\Sigma(s)$ is proportional to the rate of change of the path length $\Lambda(s)$ along the D-MORPH search trajectory. As was the case for optimization searches in Ref.~\cite{MooreChakrabarti2011}, we will show that the decrease of the slope metric $\Sigma(s)$ (i.e., the increase of the landscape ``flatness'') at an algorithmic iteration close to the landscape optimum (e.g., for $\tilde{\mathcal{D}} \approx 10^{-6}$) correlates with the increase of the search effort.

\subsection{Model control systems}
\label{sec:syst}

Various types of coupled-spin systems have been considered in TOC studies, with many works using models relevant to the liquid-state NMR, where spins are coupled via an Ising-type interaction and coupling strengths are much smaller than differences between spin frequencies \cite{Carlini2011, SchulteSporl2005, KhanejaReiss2005, KhanejaHeitmann2007}. Other models such as anisotropic controllable inter-qubit couplings \cite{Carlini2007, KoikeOkudaira2010} have also been studied. In NMR models, control fields typically address each spin separately with independent $x$ and $y$ polarizations \cite{KhanejaReiss2005, SchulteSporl2005}. 

In this work, we study TOC of generic coupled-spin model systems motivated by implementations of quantum computing in physical devices such as semiconductor quantum dots \cite{Petta2005}. In our model, each qubit has a characteristic transition frequency and is controlled by a separate field with only the $x$ polarization; exchange interactions between qubits are of the Heisenberg type with fixed isotropic coupling strengths. All system and control parameters are expressed in dimensionless units. 

Generally, we consider a system of $n$ qubits (the Hilbert space dimension is $N = 2^n$), with the model Hamiltonian of the form (\ref{eq:Ham}). The field-free Hamiltonian $H_0$ is given by
\begin{equation}
\label{ho}
H_0 = \sum_{k=1}^n \omega_k S_z^{(k)} + \sum_{k=1}^{n-1} \sum_{j=k+1}^{n} J^{(k,j)} \mathbf{S}^{(k)} \cdot \mathbf{S}^{(j)} . 
\end{equation}
Here, the operator $S_a^{(k)}$ ($a = x,y,z$) denotes the tensor product of the spin operator for the $k$th qubit with identity operators for all other qubits: 
\begin{equation}
\label{eq:Sk}
S_a^{(k)} = \underset{k-1}{\underbrace{\openone_2 \otimes \cdots \otimes \openone_2}} \otimes S_a \otimes \underset{n-k}{\underbrace{\openone_2 \otimes \cdots \otimes \openone_2}} ,
\end{equation}
where the spin operators are $\mathbf{S} = (S_x, S_y, S_z) = \frac12(\sigma_x, \sigma_y, \sigma_z)$, in terms of the Pauli matrices, and $\openone_2$ is the $2 \times 2$ identity matrix. Each qubit has a unique transition frequency $\omega_k$ (corresponding to the presence of a static magnetic field in the $z$ direction in the spin model), and isotropic coupling strengths $J^{(k,j)}$ between pairs of qubits are constant. In the simulations reported here, we used model systems with up to four qubits with frequencies $\omega_k = 20, 24, 30, 40$ and coupling constants $J^{(k,j)}$ in the range from $0.08$ to $400$. This broad range of coupling strengths represents the freedom inherent in considering coupled-spin systems in contexts other than NMR, such as semiconductor quantum dots, where interactions between qubits may be tuned by application of electric fields \cite{Petta2005, RoloffWeninPotz2009JCE, NielsenCarroll2010, GraceDominyWitzel2012}.

The control Hamiltonian $H_{\mathrm{c}}(t)$ corresponds to the application of a separate time-dependent control field polarized in the $x$ direction to each individual qubit:
\begin{equation}
\label{dipole}
 H_{\mathrm{c}}(t) = \sum_{k=1}^n \varepsilon_k(t) H_{\mathrm{c}}^{(k)} 
= \sum_{k=1}^{n} \varepsilon_k(t) S_x^{(k)} ,
\end{equation}
where $\varepsilon_k(t)$ is the control field applied to the $k$th qubit and the operator $S_x^{(k)}$ is defined by Eq.~(\ref{eq:Sk}). In the optimization procedure, each control field is labeled by the algorithmic index $s$ (see Sec.~\ref{sec:dmorph} above). The fluence of the $k$th field at the index $s$ is given by 
\begin{equation}
  \label{eq:fluence}
  f_k(s) = \int_0^T \! d t\, \varepsilon_k^2(s,t) .
\end{equation}
At the start of the optimization ($s = 0$), each field is initialized in the parameterized form:
\begin{equation}
\label{field}
\varepsilon_k(0,t) = A(t) \sum_{i=1}^{M} \sin\left(\eta_i t + \varphi_i \right) ,
\end{equation}
$t \in [0,T]$. Here, $A(t) = A_0 \exp[ -8\pi (t-T/2)^2/T^2 ]$ is the Gaussian envelope function, frequencies $\{\eta_i\}$ (corresponding to $M$ spectral components of the field; we usually use $M = 10$) are randomly selected from a uniform distribution on $[0,\Omega]$ (with $\Omega$ being the largest transition frequency in $H_0$), $\{\varphi_i\}$ are random phases on $[0,2\pi]$. The normalization constant $A_0$ is chosen so that the fluence $f_k(0)$ of each initial field is equal to 1.

The parameterized form (\ref{field}) is used only for initial control fields. At each step of the optimization algorithm after the initialization (i.e., for $s>0$), the value of the $k$th control field at each point on the time discretization mesh is allowed to vary freely and independently. This flexible set of control `knobs' $\{ \varepsilon_k(s,\Delta t), \varepsilon_k(s,2\Delta t), \ldots, \varepsilon_k(s,T) \}$ allows the fluence of the $k$th field to vary freely when $s>0$. The time step $\Delta t$ is chosen such that $\Delta t < \pi/(2\Omega)$, corresponding to the Nyquist frequency $\omega_{\mathrm{N}} = \pi / \Delta t > 2\Omega$. In agreement with the Nyquist--Shannon sampling theorem, this criterion was found to be sufficient to ensure that the time discretization does not affect the reachability of the global optimum \cite{MooreChakrabarti2011, MooreRabitz2011}. 

\subsection{Relationship between control time and robustness to additive white control noise}
\label{sec:robustness}

An important motivation for finding the minimum time necessary to enact a target unitary transformation is to improve the robustness of the gate operation to noise in control fields. Specifically, additive white noise (AWN) in optimal control fields induces gate errors that are linearly proportional to $T$ \cite{Brif:robustness}; a brief outline of this analysis is provided here. In the presence of additive noise, the actual control field is given by $\varepsilon(t) + \xi(t)$, where $\xi(t)$ is a classical stochastic variable. For white noise, $\xi(t)$ has zero mean: $\mathbb{E}\{\xi(t)\} = 0$, and is delta-correlated: $\mathbb{E}\{\xi(t) \xi(t')\} = \sigma^2 \delta(t-t')$. Here, $\mathbb{E}\{ \cdot \}$ denotes the statistical expectation value over all noise realizations and $\sigma^2$ is the variance of the noise amplitude distribution. For AWN in multiple control fields, in general, we should also consider cross-correlations: $\mathbb{E}\{\xi_k(t) \xi_j(t')\} = \sigma^2 \beta_{k j} \delta(t-t')$, where $0 \leq \beta_{k j} \leq 1$ and $\beta_{k k} = 1$ (if noise processes in different control fields are independent, then $\beta_{k j} = \delta_{k j}$).

According to the analysis in Ref.~\cite{Brif:robustness}, for weak AWN in optimal control fields $\{\varepsilon_k^{\star}(t)\}$, the statistical expectation value of the normalized distance $\tilde{\mathcal{D}}$ is approximated (by expanding up to the second order in the noise amplitude) as
\begin{equation}
\label{noise-1}
 \mathbb{E}\{ \tilde{\mathcal{D}} \} \approx \frac{1}{2} \sigma^2 \sum_{k,j = 1}^n \beta_{k j} 
\int_0^T d t\, \tilde{\mathsf{H}}_{k j}^{\star} (t,t) .
\end{equation}
Here, $\tilde{\mathsf{H}}^{\star}(t,t')$ denotes the Hessian matrix of $\tilde{\mathcal{D}}$ evaluated at the optimum, and the diagonal elements (i.e., for $t = t'$) of its blocks are time-independent \cite{Brif:robustness, HoDominyRabitz2009PRA}:
\begin{equation}
  \label{eq:Hess-kj}
  \tilde{\mathsf{H}}_{k j}^{\star} (t,t) = \left.
\frac{\delta^2 \tilde{\mathcal{D}}}{\delta \varepsilon_k(t) \delta \varepsilon_j(t)}
\right|_{\varepsilon^{\star}} 
= \frac{1}{2 N} \mathrm{Tr}\left[ H_{\mathrm{c}}^{(k)} H_{\mathrm{c}}^{(j)} \right] . 
\end{equation}
Substituting Eq.~(\ref{eq:Hess-kj}) into Eq.~(\ref{noise-1}), one obtains an expression that reveals the linear dependence of the expected gate error on $T$:
\begin{equation}
\label{noise-2}
 \mathbb{E}\{ \tilde{\mathcal{D}} \} \approx \frac{1}{4 N} \sigma^2 T \sum_{k,j = 1}^n \beta_{k j} 
 \mathrm{Tr}\left[ H_{\mathrm{c}}^{(k)} H_{\mathrm{c}}^{(j)} \right] . 
\end{equation}
For the control Hamiltonian of Eq.~(\ref{dipole}) employed here, $ \mathrm{Tr}[ H_{\mathrm{c}}^{(k)} H_{\mathrm{c}}^{(j)} ] = \mathrm{Tr}[ S_x^{(k)} S_x^{(j)} ] = (N/4) \delta_{k j}$. Using this result (together with $\beta_{k k} = 1$) in Eq.~(\ref{noise-2}), we finally obtain:
\begin{equation}
\label{nsc}
  \mathbb{E}\{ \tilde{\mathcal{D}} \} \approx \frac{1}{16}\sigma^2 n T.
\end{equation}
Equation~(\ref{nsc}) shows that the error in the objective value is expected to grow linearly both in the number of qubits $n$ and control time $T$. Thus, it is practically important to minimize $T$ when AWN in control fields is present. 

\subsection{PFT algorithm for quantifying the trade-off between minimizations of distance and control time}
\label{sec:pft}

The PFT algorithm introduced here is designed to identify the Pareto front for the dual objectives of minimizing $\tilde{\mathcal{D}}$ and $T$, as well as to explore the corresponding domain of the optimal control landscape. We use PFT to move along the Pareto front by identifying optimal control solutions corresponding to different values of $T$. Analogous to the algorithmic index $s$ describing the progression of a D-MORPH search, we define the indexing variable $p$ to describe the progress of the PFT algorithm. The PFT algorithm works as follows: 
\begin{enumerate}
 \item Select a starting value of the control time, $T_0 = T(p=0)$. Then run the D-MORPH optimization (as described in Sec.~\ref{sec:dmorph}), starting from a set of randomly selected initial control fields $\{\varepsilon_k(p=0,s=0,t)\}$, until it converges to a set of optimal fields $\{\varepsilon_k(p=0,s=s^{\star},t)\}$ that minimizes $\tilde{\mathcal{D}}$.
 \item Reduce the value of the control time $T$, so that $T(p+1) = T(p) - \Delta T$, where $\Delta T$ is an increment on the order of $\Delta T \lesssim 0.01 T$.
 \item Resample each of the optimal control fields in the set $\{\varepsilon_k(p,s^\star,t)\}$ on the updated time interval $[0, T(p+1)]$. Employ the resulting set of fields as the initial guess $\{\varepsilon_k(p+1,s = 0,t)\}$ for the next D-MORPH optimization that proceeds to identify the next set of optimal fields $\{\varepsilon_k(p+1,s=s^{\star},t)\}$.
\end{enumerate}
Steps 2 and 3 are repeated until the D-MORPH optimization can no longer attain the desired value of $\tilde{\mathcal{D}}$. Specifically, we set the PFT ``stop value'' to $\tilde{\mathcal{D}} = 10^{-2}$ when the goal is to explore the ``competitive'' part of the Pareto front, and to $\tilde{\mathcal{D}} = 10^{-8}$ when the goal is to only determine the critical time $T^{\ast}$. The process of decreasing $T$ in small increments coupled with the use of optimal control fields as the initial guess for the D-MORPH run in the next PFT iteration biases towards identifying families of related control solutions along the Pareto front. Therefore, running multiple PFT trajectories beginning from different random initial control fields at various values of $T_0$ may be useful for proper identification of the Pareto front.

\section{Exploring the Pareto front for control of two-qubit gates}
\label{sec:2q}

In this section, we explore the Pareto front for fidelity- and time-optimal control of unitary transformations in two-qubit systems. In Sec.~\ref{sec:ex}, we study in detail how the optimization search effort, landscape metrics, and optimal control fields change along the Pareto front for the controlled NOT (CNOT) target gate and one representative set of system parameters. In Secs.~\ref{sec:2q2} and \ref{sec:2q3}, we identify Pareto fronts and determine critical times for a variety of target gates and inter-qubit coupling strengths.

\subsection{Optimization search effort and Pareto front exploration for CNOT gate}
\label{sec:ex}

As an illustrative case on which to conduct a detailed examination of the search effort dependence on the control time and properties of the distance-time Pareto front, we consider the objective of performing the CNOT gate in the two-qubit model system ($n = 2$) with $\omega_1 = 20$, $\omega_2 = 24$, and $J^{(1,2)} = 0.8$. Since the system Hamiltonian is traceless, the target transformation $W$ is defined as the CNOT gate with a global phase factor chosen so that $W$ is in SU(4):
\begin{equation}
\label{cnot}
 W_{\mathrm{CNOT}} = e^{-i \pi/4} \begin{pmatrix}
           1& 0& 0 &0\\
	       0& 1 &0& 0\\
	       0& 0& 0& 1\\
	       0& 0& 1 &0
              \end{pmatrix}.
\end{equation}

\begin{figure}[hb]
 \includegraphics[width=7.7cm]{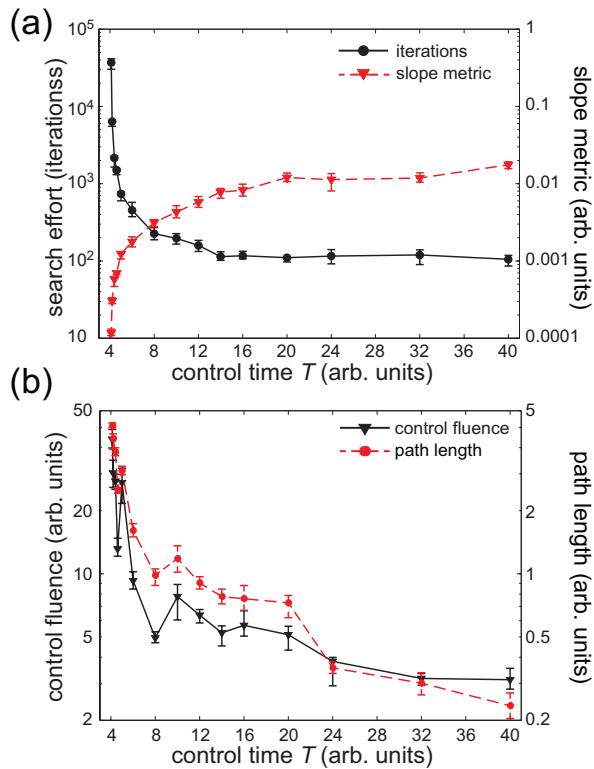}
\caption{(Color online) The search effort and metrics of the control landscape as functions of the control time $T$ for D-MORPH optimizations performed at select fixed values of $T$. The target gate is $W_{\mathrm{CNOT}}$. (a) Search effort (black circles with solid line, left-side ordinate) and slope metric (the $L_2$-norm of the landscape gradient) $\Sigma(s)$ evaluated at the $s$ value corresponding to $\tilde{\mathcal{D}} \approx 10^{-6}$ (red triangles with dashed line, right-side ordinate). (b) Total fluence $f^{\star}$ of optimal control fields (black triangles with solid line, left-side ordinate) and total path length $\Lambda^{\star}$ along the search trajectory (red circles with dashed line, right-side ordinate). Circles and triangles show average values and error bars denote the left and right standard deviation over the sample of 10 D-MORPH searches with randomly selected initial control fields performed for each value of $T$.}
\label{fig:itergrad}
\end{figure}

\subsubsection{Search effort of D-MORPH optimizations and its relationship to landscape metrics}
\label{sec:dm}

For the selected target gate and control system, the critical time value $T^{\ast} \approx 4.12$ was found by running PFT trajectories (see Sec.~\ref{sec:pf} below for details). In order to study how the control time affects the optimization search effort, D-MORPH trajectories were obtained for 15 values of $T$ between $T = 40$ and $T = 4.12$. For each value of $T$, 10 D-MORPH optimization runs beginning from different random initial fields of unit fluence were performed. All optimizations reached the desired objective value $\tilde{\mathcal{D}} \leq 10^{-8}$ and no trapping or slowdown of searches at a suboptimal distance value was observed (including runs for $T = T^{\ast}$). The search effort, however, increased as $T$ was made smaller, especially for $T \leq 6$. This behavior is shown in Fig.~\ref{fig:itergrad}(a) as a plot of the number of D-MORPH algorithmic iterations (averaged over 10 searches started from random initial fields) versus $T$. In particular, for $T \leq 6$, the search effort increases superexponentially (note the logarithmic scale of the left-side ordinate) and is well approximated as $\exp(a T^{-b} + c)$, where $a \approx 2.1 \times 10^7$, $b \approx 11.0$, and $c \approx 6.1$. The corresponding decrease in the slope metric $\Sigma(s)$ evaluated at the $s$ value corresponding to $\tilde{\mathcal{D}} \approx 10^{-6}$ is also shown in Fig.~\ref{fig:itergrad}(a) (with values on the right-side ordinate). The complementary trends for the search effort and gradient norm indicate that the control landscape in regions near the optimum becomes ``flatter'' as $T$ decreases.

To examine how the distance between the initial and optimal fields grows as $T$ decreases towards $T^{\ast}$, we consider the total fluence of optimal control fields, $f^{\star} = \sum_{k} f_k(s^{\star})$, and the total path length $\Lambda^{\star} = \Lambda(s^{\star})$ traveled along the search trajectory to reach the optimum. These quantities (averaged over 10 searches started from random initial fields) are plotted versus $T$ in Fig.~\ref{fig:itergrad}(b). Both $f^{\star}$ and $\Lambda^{\star}$ rise as $T$ decreases, and this rise significantly accelerates for $T \leq 8$. Since all D-MORPH searches began from unit-fluence fields, the value of $f^{\star}$ is an indicator of the distance between the initial and optimal fields, which explains the similarity in the trends of $f^{\star}$ and $\Lambda^{\star}$. The behavior of the fluence as a function of the control time can be qualitatively explained by a simple example of a two-level system driven by a resonant control field. The rotation angle produced by the control Hamiltonian on the Bloch sphere during time $T$ is $\Omega_R T$, where $\Omega_R$ is the Rabi frequency which is proportional to the control-field amplitude $\varepsilon_0$. If the goal is to generate the same rotation as $T$ changes, the optimal-field amplitude should scale as $\varepsilon^{\star}_0 \propto 1/T$. Since the fluence can be approximated as $f \sim \frac{1}{2} \varepsilon_0^2 T$, the optimal-field fluence scales as $f^{\star} \propto 1/T$. Of course, for a multi-qubit system controlled by several external fields, the dynamics is much more complicated (in fact, even for a single qubit, $1/T$ scaling is periodically modulated by the effect of free evolution). Nevertheless, this simple picture helps to explain qualitatively why the fluence rises when $T$ decreases. The non-monotonicity of $f^{\star}$ and $\Lambda^{\star}$ as functions of $T$, seen in Fig.~\ref{fig:itergrad}(b), is explained by the fact that, as $T$ changes, free evolution takes the system closer to or further away from the target, thus modulating the amount of control-field energy required to reach the optimum.

\begin{figure}[htbp]
 \includegraphics[width=7.7cm]{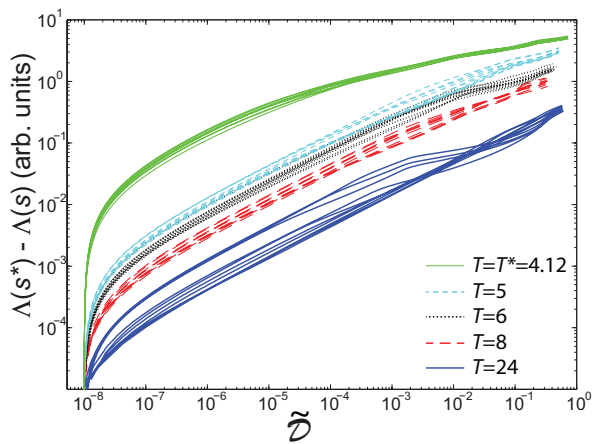}
\caption{(Color online) Path length difference $\Lambda(s^{\star})-\Lambda(s)$ along the optimization trajectory as a function of the objective $\tilde{\mathcal{D}}$. As the optimization progresses, the path length difference decreases towards zero along with $\tilde{\mathcal D}$. The target gate is $W_{\mathrm{CNOT}}$. Each value of the control time $T$ is denoted by line style and color in the legend. The order of trajectories in the legend corresponds to that on the figure, with the $T^{\ast}$ trajectories (green, solid line) at the top, and $T$ increasing from top to bottom. Different trajectories for the same value of $T$ correspond to 10 D-MORPH searches with randomly selected initial control fields.}
\label{fig:pathf}
\end{figure}

Further insight into the dependence of the control landscape structure on the control time can be gained by examining the path length $\Lambda(s)$ accumulated as the D-MORPH search progresses. The value of $\Lambda$ increases monotonically along the search trajectory from $s = 0$ to $s = s^{\star}$; correspondingly, the path length difference $\Lambda(s^{\star})-\Lambda(s)$ indicates the extent to which the control field has reached its optimal form. Figure~\ref{fig:pathf} shows $\Lambda(s^{\star})-\Lambda(s)$ as a function of the objective value $\tilde{\mathcal{D}}$ for optimization trajectories with selected values of $T$, with 10 trajectories corresponding to different random initial fields shown for each $T$. This plot illustrates the dependence of the search trajectory on $T$. For all searches with $T > T^{\ast}$, the path length difference follows a similar power law (which appears as a linear change on the log--log plot) over the range from $\tilde{\mathcal D} \sim 10^{-3}$ to $\tilde{\mathcal D} \sim 10^{-7}$. In contrast, the searches with $T = T^{\ast}$ approach the final path length $\Lambda(s^{\star})$ much more slowly until $\tilde{\mathcal D}$ becomes very close to the optimum; there, the path length changes very quickly, as $\Lambda(s^{\star})-\Lambda(s)$ drops by three orders of magnitude between $\tilde{\mathcal D} \simeq 2 \times10^{-8}$ and $\tilde{\mathcal D} \simeq 10^{-8}$. This result shows that at $T^{\ast}$, large changes of the control field occur very near the optimum.

\subsubsection{PFT results}
\label{sec:pf}

\begin{figure}[t]
\includegraphics[width=7.7cm]{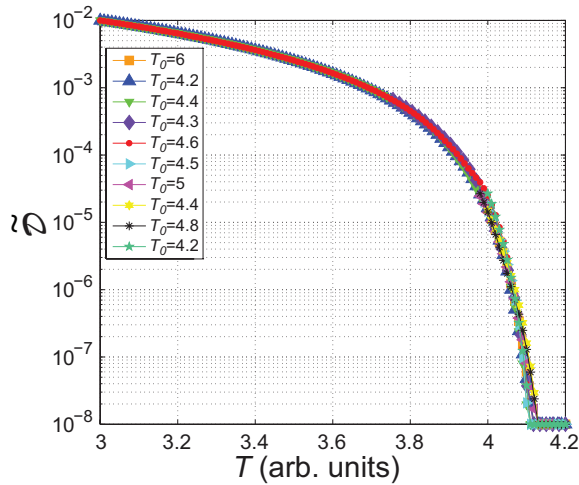}
 \caption{(Color online) The normalized distance $\tilde{\mathcal{D}}$ plotted versus the control time $T$ for 10 separate PFT trajectories denoted by shape and color. The target gate is $W_{\mathrm{CNOT}}$. The PFT trajectories were started at different values of $T_0$ (given in the legend) with different randomly selected initial control fields, but are very closely aligned, which suggests the existence of a unique Pareto front.}
\label{fig:pareto1}
\end{figure}

The PFT procedure described in Sec.~\ref{sec:pft} was performed for the purposes of (a) identifying the critical time $T^{\ast}$ (determined numerically as the minimum control time at which $\tilde{\mathcal{D}}\leq 10^{-8}$ can be achieved) and (b) following the Pareto front to obtain the best attainable $\tilde{\mathcal{D}}$ value as a function of $T$ in the ``competitive'' region of $T < T^{\ast}$. A total of $10$ PFT trajectories beginning from random initial control fields at different values of $T_0$ (ranging from $T_0 = 4.2$ through $T_0 = 6$) were generated for the target gate $W_{\mathrm{CNOT}}$ and the same two-qubit system as used in the D-MORPH optimizations in Sec.~\ref{sec:dm} above. Four of these trajectories were run along the Pareto front by decreasing $T$ until it was impossible to attain $\tilde{\mathcal{D}} \leq 10^{-2}$, while the remaining trajectories were run only to values of $T$ ranging from $T = 3.8$ to $T = 4$ due to computational expense. All 10 trajectories, shown in Fig.~\ref{fig:pareto1}, are very closely aligned for all values of $T$, with the critical time value estimated as $T^{\ast} = 4.12 \pm 0.01$ (i.e., the dispersion of the $T^{\ast}$ value is on the order of the PFT step $\Delta T$). This near coincidence of different PFT trajectories indicates that, for a given quantum system and a target gate, there exists a unique distance-time Pareto front. Correspondingly, the minimum control time needed to enact a target unitary transformation with a high fidelity appears to be an inherent property of the controlled quantum system, independent of the path taken to identify $T^{\ast}$.

The Pareto front for $T \lesssim T^{\ast}$ has the unfavorable property of an extremely steep slope, i.e., a small decrease in $T$ below $T^{\ast}$ results in a very large increase in $\tilde{\mathcal{D}}$. Specifically, $\tilde{\mathcal{D}}$ rises more than three orders of magnitude from $\tilde{\mathcal{D}} < 10^{-8}$ to $\tilde{\mathcal{D}} > 10^{-5}$ with a relatively small decrease in $T$ from $T = 4.12$ to $T = 4.0$. Beyond this steep region, the distance growth moderates with decreasing $T$, such that $\tilde{\mathcal{D}} \approx 10^{-2}$ can still be obtained at $T = 3.0$. However, since a fault-tolerant quantum computation requires very low gate error rates (typically, less than $10^{-4}$) \cite{NielsenChuang2000, Gottesman2009, *Divincenzo2009}, the steep slope of the Pareto front immediately below $T^{\ast}$ presents a fundamental limitation on gate implementation times. Furthermore, uncertainty introduced under experimental conditions would make operation near $T^{\ast}$ difficult because small errors in control time could cause substantial decreases in attainable fidelity.

\begin{figure*}[htbp]
\includegraphics[width=15.8cm]{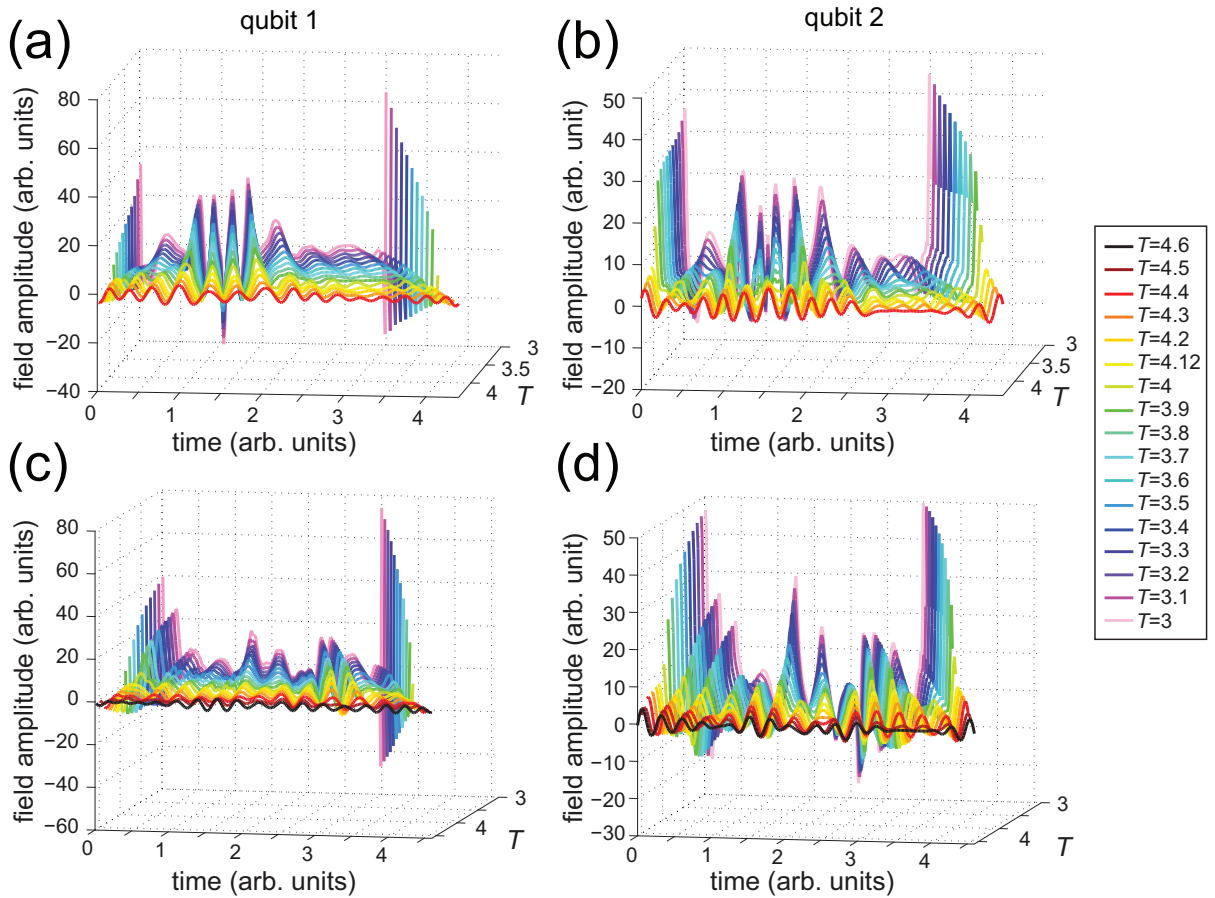}
 \caption{(Color online) Optimal control fields $\{ \varepsilon_k^{\star}(t) \}$ for the target gate $W_{\mathrm{CNOT}}$. The control fields for the first qubit ($k = 1$) are shown in (a) and (c), and the control fields for the second qubit ($k = 2$) in (b) and (d). These plots show fields obtained for various values of the control time $T$, ranging from $T = T_0$ (the starting point of the PFT trajectory) to $T = 3$ (where the best attainable objective value is $\tilde{\mathcal{D}} \approx 0.01$). Each value of $T$ is denoted by color in the legend, with yellow (lightest gray) corresponding to $T = T^{\ast} = 4.12$. The shown sets of fields represent two PFT trajectories started at $T_{0} = 4.4$ ((a) and (b)) and $T_{0} = 4.6$ ((c) and (d)); these PFT trajectories are presented among several others in Fig.~\ref{fig:pareto1}. An apparent increase in the DC field component (i.e., a positive shift of the entire field) as $T$ decreases is an artifact of the perspective of the three-dimensional plot; the field power spectra (not shown) do not reveal a significant zero-frequency component.}
\label{fig:fieldcnot}
\end{figure*}

In addition to being a reliable numerical method to identify the distance-time Pareto front (and, in particular, determine the value of $T^{\ast}$), the PFT algorithm generates families of related control fields that minimize the objective $\tilde{\mathcal{D}}$ for control times within a selected interval. While all PFT trajectories achieve essentially the same minimum value of $\tilde{\mathcal{D}}$ at a given $T$, each trajectory produces a distinct family of optimal control solutions. The ``evolution'' of optimal control fields within such a family, corresponding to the change of $T$ along a PFT trajectory, is visualized in Fig.~\ref{fig:fieldcnot}. Specifically, two families of optimal fields obtained for PFT trajectories initialized at $T_0 = 4.4$ and $T_0 = 4.6$ are shown in Figs.~\ref{fig:fieldcnot}(a), (b) and \ref{fig:fieldcnot}(c), (d), respectively. For each of these two families, pairs of fields (corresponding to separate control fields acting on two qubits) are shown for several values of $T$ ranging from $T = T_0$ to $T = 3.0$. 

The optimal control fields obtained for different PFT trajectories have distinct shapes, which is expected based on the existence of an infinite number of optimal solutions \cite{DemiralpRabitz1993}. Nevertheless, the fields share a number of common features. First, field amplitudes increase as $T$ decreases. As explained in Sec.~\ref{sec:dm}, this amplitude behavior is needed to maintain the required rotation angle as $T$ changes, and the associated change of the field fluence roughly follows $1/T$ scaling. Second, for $T < T^{\ast}$, each field exhibits spikes at $t = 0$ and $t = T$ that grow in amplitude as $T$ decreases. For $T < 3.9$, these spikes are the largest amplitude features of the control fields. Besides the two families shown in Fig.~\ref{fig:fieldcnot}, this field feature was observed for all other PFT trajectories that were run below $T^{\ast}$. Similar characteristics of optimal fields were also observed for TOC of a three-qubit system in Ref.~\cite{SchulteSporl2005}. We made no attempt to impose any control constraints that would suppress such spiky features; the extreme difficulty of the distance minimization in the region $T < T^{\ast}$ indicates that further constraints could adversely affect attainable values of $\tilde{\mathcal{D}}$ and thus preclude reaching the genuine distance-time Pareto front.

The results presented in this section suggest that the location of the Pareto front is essentially independent of the PFT trajectory taken for the present numerical model. Thus, for other selections of the target gate and/or system parameters considered in Secs.~\ref{sec:2q2} and \ref{sec:2q3} below, Pareto fronts were identified by running only one complete (i.e., followed until $\tilde{\mathcal{D}} \leq 10^{-2}$ becomes unattainable) PFT trajectory. For cases where only identification of $T^{\ast}$ was desired, the PFT algorithm was stopped soon after $\tilde{\mathcal{D}} \leq 10^{-8}$ was no longer attainable, and at least three trajectories started with different initial random fields were run in this fashion in order to verify the obtained value of $T^{\ast}$. 

\subsubsection{Efficiency of the PFT algorithm}

In Sec.~\ref{sec:dm}, a superexponential increase of the optimization search effort (in terms of algorithmic iterations) was observed as the control time $T$ decreases and approaches $T^{\ast}$. The D-MORPH searches considered in Sec.~\ref{sec:dm} were initialized at randomly selected fields for all values of $T$, with typical initial objective values $\tilde{\mathcal{D}}(s=0) \sim 0.5$. In contrast, the PFT algorithm employs random initial fields only for the search with the starting value of the control time, $T_0 = T(p=0)$; every consequent search with $T(p+1) < T_0$ is initialized at the fields $\{\varepsilon_k(p,s^{\star},t)\}$ that are optimal for the preceding search with $T(p) = T(p+1) + \Delta T$. Correspondingly, D-MORPH searches along a PFT trajectory begin, for $p > 0$, with initial objective values $\tilde{\mathcal{D}}(s=0) \sim 0.01$ to $0.05$ and thus may be expected to reach $\tilde{\mathcal{D}} \leq 10^{-8}$ with a smaller number of algorithmic iterations. Here, we investigate to what degree the PFT algorithm can lower the search effort, as compared to D-MORPH optimizations with random initial fields.

\begin{figure}[t]
\includegraphics[width=7.7cm]{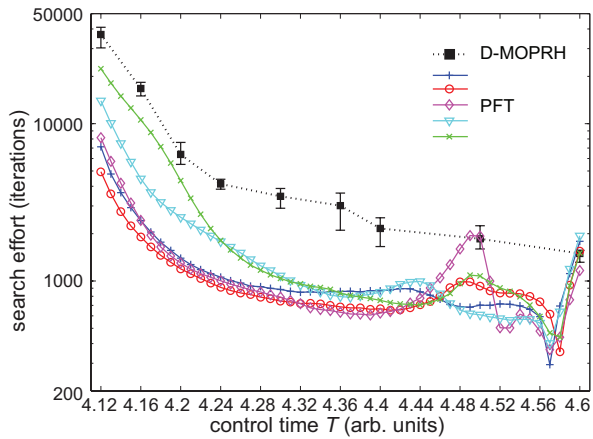}
\caption{(Color online) The search effort versus the control time $T$ for five PFT trajectories (denoted by distinct symbol shapes and colors) and D-MORPH searches with random initial fields (black squares with error bars denoting the average and standard deviation, respectively, over the set of 10 runs). At each value of $T \leq 4.45$, the PFT search effort is sizably smaller than the average D-MORPH effort.}
\label{fig:dmorphpftcompare}
\end{figure}

To directly compare the two methods, we ran (a) five PFT trajectories from $T_0 = 4.6$ to $T^{\ast} = 4.12$ and (b) sets of 10 D-MORPH searches with random initial fields at selected values of $T \in [4.12, 4.6]$. Figure~\ref{fig:dmorphpftcompare} presents the search effort as a function of $T$ for each PFT trajectory (colored circles, diamonds, triangles, crosses, and x's) and for each set of D-MORPH searches (black squares and error bars indicating the average and standard deviation, respectively, over the set of 10 runs). The results show that, while different PFT trajectories exhibit varying convergence speeds, at each value of $T \leq 4.45$, the  search effort for PFT is significantly lower than that for randomly initialized D-MORPH, typically by a factor of 2 to 4. At $T^{\ast} = 4.12$, the convergence of the best PFT trajectory was faster than that of the average D-MORPH search by a factor of $\sim 8$; even the worst PFT trajectory outperformed the average D-MORPH search by a factor of $\sim 2$. Thus, the search effort can be substantially lowered if a D-MORPH search is initialized at control fields that are optimal for a related search (here, with a slightly larger value of $T$) rather than at random fields.

It is also of interest to compare the PFT algorithm to methods that aim at minimizing $T$ by including penalty terms into the cost functional  \cite{MishimaYamashita2009a, *MishimaYamashita2009b, Lapert2012PRA}. For the latter approach, a recent work \cite{Lapert2012PRA} assigned \emph{a priori} weights to a term in the cost functional that penalizes the control field fluence and control time. It is known \cite{RajWu2008b} that OCT algorithms employing such cost functionals are typically incapable of identifying the genuine Pareto front. Indeed, the algorithm used in Ref.~\cite{Lapert2012PRA} converges to a value of $T$ that overestimates the actual critical time, while also underestimating the achievable gate fidelity. A PFT trajectory that we ran for the quantum system and target gate used in Ref.~\cite{Lapert2012PRA} identified $T^{\ast} \approx 1.80$ with $\mathcal{F} > 1 - 10^{-8}$, while Ref.~\cite{Lapert2012PRA} reported convergence to $T \approx 2.01$ with $\mathcal{F} \approx 1 - 7 \times10^{-6}$. These results suggest that the PFT algorithm is better suitable for accurately determining the fidelity-time Pareto front and, in particular, the true value of $T^{\ast}$ than methods that employ time-penalizing terms in the cost functional.

\subsection{Pareto fronts and critical times for various target gates}
\label{sec:2q2}

Simulations presented in this section examine how the distance-time Pareto front and, in particular, the critical time are affected by the choice of the target transformation $W$ for the same two-qubit system as considered in Sec.~\ref{sec:ex}. Pareto fronts and values of $T^{\ast}$ for various target gates are identified using the PFT procedure, as described in detail above. In addition to the $W_{\mathrm{CNOT}}$ gate of Eq.~(\ref{cnot}), we considered a number of gates, all incorporating appropriate global phase factors to make them elements of SU($N$). The SWAP gate has the form
\begin{align}
W_{\mathrm{SWAP}} = e^{-i \pi/4}\begin{pmatrix}
1&0&0&0\\
0&0&1&0\\
0&1&0&0\\
0&0&0&1\\
\end{pmatrix}.\label{swap}
\end{align}
We also consider the square root of SWAP (an entangling two-qubit gate), given by
\begin{align}
W_{\sqrt{\mathrm{SWAP}}} = e^{i \pi/8}\begin{pmatrix}
1&0&0&0\\
0&\frac{1}{\sqrt{2}}e^{i \pi/4}&\frac{1}{\sqrt{2}}e^{-i \pi/4}&0\\
0&\frac{1}{\sqrt{2}}e^{-i \pi/4}&\frac{1}{\sqrt{2}}e^{i \pi/4}&0\\
0&0&0&1\\
\end{pmatrix}.\label{sqrtswap}
\end{align}
Matrix elements of the Quantum Fourier Transform (QFT) gate generalized for $n$ qubits are given by
\begin{equation}
\label{qft}
W_{\mathrm{QFT},n}(j,k) = \frac{1}{\sqrt{N}} e^{5 i \pi/(2 N)} \omega^{j k} ,
\end{equation}
where $j,k = 0,1,\ldots, N-1$ and $\omega = e^{2 i \pi/N}$ (recall that $N = 2^n$). For two-qubit systems ($n = 2$), the QFT gate is
\begin{align}
W_{\mathrm{QFT},2} = \frac{1}{2}e^{5 i \pi/8}\begin{pmatrix}
1&1&1&1\\
1&\omega&\omega^2&\omega^3\\
1&\omega^2&\omega^4&\omega^6\\
1&\omega^3&\omega^6&\omega^9
\end{pmatrix}.\label{qft2}
\end{align}
We also consider a similar gate denoted as $W_{\mathrm{QFT}',n}$, which is given by Eq.~(\ref{qft}) with $j,k = 1,2,\ldots, N$. Finally, we consider controlled phase (CPHASE) gates, of the form
\begin{align}
W_{\mathrm{CPHASE}}(\alpha) = e^{-i \alpha/4}\begin{pmatrix}
1&0&0&0\\
0&1&0&0\\
0&0&1&0\\
0&0&0&e^{i \alpha}\\
\end{pmatrix},\label{cphase}
\end{align}
with the phases $\alpha = \pi$ and $\alpha = \pi/2$.

In addition to the explicit forms of the CNOT, SWAP, $\sqrt{\mathrm{SWAP}}$, QFT, QFT$'$, and CPHASE gates presented above, we also consider the effect of changing the global phase of the target gate. In earlier work by Schulte-Herbr\"uggen \emph{et al.}~\cite{SchulteSporl2005}, different values of $T^{\ast}$ were found for distinct global phases of the three-qubit QFT gate in a linear spin-chain system with Ising-type interactions. Identifying critical times that correspond to all possible values of the target gate's global phase is important in the context of TOC. For an $n$-qubit gate $W \in \mathrm{SU}(N)$, transformations with $N$ distinct global phase values are allowed:
\begin{eqnarray}
&& W(\phi_m) = e^{i \phi_m} W, \nonumber \\
&& \phi_m = 2 m \pi/N, \quad m = 0,1,\ldots,N-1.
\label{phi-m}
\end{eqnarray}
For two-qubit gates  considered in this section, four distinct global phase values are allowed in SU(4): $\phi_m = m \pi/2$, $m = 0,1,2,3$. The critical time for the transformation $W(\phi_m)$ is denoted as $T^{\ast}(\phi_m)$.

\begin{figure}[t]
\includegraphics[width=7.7cm]{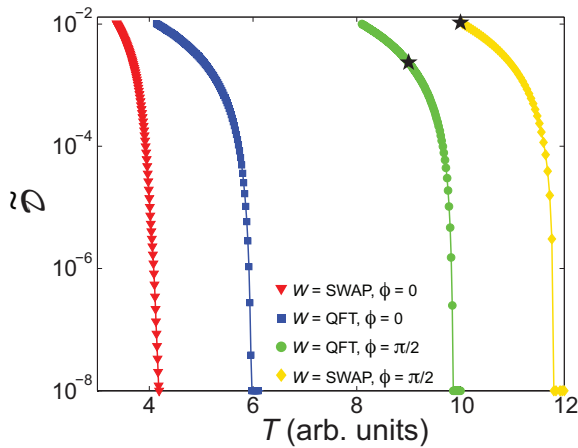}
 \caption{(Color online) Distance-time Pareto fronts for the QFT and SWAP gates with global phases $\phi = 0$ and $\phi = \pi/2$. Each front is denoted by a distinct symbol shape and color, as listed in the legend. Fronts for $\phi = \pi$ and $\phi = 3\pi/2$ (not shown here) closely match those for $\phi = 0$ and $\phi = \pi/2$, respectively. The significant difference in front locations (and corresponding $T^{\ast}$ values) demonstrates the importance of the gate's global phase in TOC. The black stars denote the average values of $\tilde{\mathcal{D}}$ obtained from the searches reported in Table~\ref{t:nophase}.}
\label{fig:paretophi}
\end{figure}

The effect of global phase on the location of the distance-time Pareto front is demonstrated in Fig.~\ref{fig:paretophi} for the SWAP and QFT gates with $\phi = 0$ and $\phi = \pi/2$. It is seen that the Pareto fronts for $\phi = 0$ are shifted significantly to the left on the time axis (and have correspondingly smaller values of $T^{\ast}$) as compared to the fronts for $\phi = \pi/2$. This behavior shows that the global phase is an important parameter in TOC of unitary transformations, consistent with the findings of Ref.~\cite{SchulteSporl2005}. We also observed for all target gates considered here that the Pareto front for $\phi = \pi$ lies very close to the one for $\phi = 0$, and the Pareto front for $\phi = 3\pi/2$ lies very close to the one for $\phi = \pi/2$. The pairwise separation of $T^{\ast}(\phi_m)$ values for global phases with $m = \{0,2\}$ and $m = \{1,3\}$ is reported in Table~\ref{t:phi} for several target transformations. For all gates except CNOT, there is a significant difference in the $T^{\ast}$ values corresponding to the two pairs of $\phi_m$ values.

\begin{table}[htbp]
\caption{\label{t:phi}Critical times $T^{\ast}(\phi_m)$ obtained using the PFT method for the selected set of target gates and four global phase values $\phi_m = m\pi/2$ ($m = 0, 1, 2, 3$).}
\begin{ruledtabular}
\begin{tabular}{crrrr}
Gate &\multicolumn{4}{c}{$T^{\ast}(\phi_m)$} \\
 & $m = 0$ & $m = 1$ & $m = 2$ & $m = 3$ \\
\hline
CNOT & 4.12 & 4.11 & 4.15 & 4.09 \\
SWAP & 4.19 & 11.80 & 4.34 & 11.84 \\
$\sqrt{\mathrm{SWAP}}$ & 2.26 & 9.84 & 2.20 & 9.82 \\
QFT & 5.98 & 9.86 & 5.94 & 9.86 \\
QFT$'$ & 6.19 & 9.92 & 5.96 & 9.90 \\
CPHASE($\pi$) & 4.22 & 3.98 & 4.21 & 3.96 \\
CPHASE($\pi/2$) & 2.25 & 5.98 & 2.38 & 5.96 \\
\end{tabular}
\end{ruledtabular}
\end{table}

The dependence of the critical time $T^{\ast}$ on the gate's global phase $\phi$ has important practical implications for optimizations in numerical simulations and, potentially, experiments, where implementing the target gate in minimum possible time is often desirable. In a common situation where initial control fields are selected randomly, a search aimed at minimizing the distance $\tilde{\mathcal{D}}$ for the target gate with a certain value of $\phi$ may, in principle, converge to a control solution that enacts the gate with another value of $\phi$. For the target transformation in SU($N$), the discreteness of allowed phase values $\phi_m$ of Eq.~(\ref{phi-m}) means that the probability of such an event depends on the distance between transformations with nearby values of $\phi$. While the fraction of searches that converge to solutions corresponding to a ``wrong'' value of $\phi$ is negligible for two-qubit gates, it will increase for multi-qubit gates as the number $N$ of allowed phase values grows (see Sec.~\ref{sec:multiq} below for more details). Furthermore, if the phase-independent distance $\mathcal{G}$ of Eq.~(\ref{ind}) is employed (which is likely in practical situations where the global phase of a  gate is undetectable), randomly initialized searches will converge to control solutions corresponding to all $\phi_m$ values (the probability for each $\phi_m$ should be equal to $1/N$, provided that a local search algorithm is employed and initial control fields are generated in a truly random manner). Therefore, in order to make all transformations $W(\phi_m)$ reachable, the control time $T$ must be not smaller than any of the critical times $T^{\ast}(\phi_m)$, i.e., $T \geq \max \{ T^{\ast}(\phi_m) \}$. 

In particular, due to the difference between the $T^{\ast}(\phi_m)$ values, an undesirable situation will arise if $T$ is selected such that $T^{\ast} (\phi_m) < T < T^{\ast} (\phi_{m'})$. Then $W(\phi_m)$ will be reachable, while $W(\phi_{m'})$ will not, and therefore local searches converging to $W(\phi_m)$ will attain a desired objective value (e.g., $\mathcal{G} \leq 10^{-8}$), while local searches moving towards $W(\phi_{m'})$ will be unable to improve the objective value beyond the limit set by the corresponding Pareto front. If such a situation is encountered in a numerical simulation, it may appear as if a fraction of searches are ``trapped,'' which seems to contradict the trap-free topology of the quantum control landscape. However, in reality, this behavior arises due to the limitation on the control time (i.e., a constraint on a critical control resource) that makes one or more of the target transformations unreachable. Obviously, if the control time is increased to satisfy the condition $T \geq \max \{ T^{\ast}(\phi_m) \}$, this spurious ``trapping'' will be completely eliminated.

We explored this effect by running 500 D-MORPH searches (with random initial fields), aimed at minimizing the phase-independent distance $\mathcal{G}$ for QFT and SWAP gates with $T = 9$ and $T = 10$, respectively. These searches may approach solutions corresponding to any $\phi_m$ ($0, \pi/2, \pi, 3\pi/2$), with optimization trajectories determined by initial fields. As expected based on the $T^{\ast}$ values in Table~\ref{t:phi}, all searches that converged to $W(\phi = 0)$ and $W(\phi = \pi)$ attained desired objective values $\mathcal{G} \leq 10^{-8}$, while all searches that moved towards $W(\phi = \pi/2)$ and $W(\phi = 3\pi/2)$ halted well before reaching the target. Table~\ref{t:nophase} presents the percentage of searches approaching each target transformation $W(\phi_m)$, along with the associated mean value of $\mathcal{G}$. For the ``trapped'' cases (occurring for $\phi = \pi/2$ and $\phi = 3\pi/2$), the standard deviation around the mean objective value $\overline{\mathcal{G}}$ is less than 0.5$\%$, which indicates that all these runs failed to reach the target due to the same cause. For these searches, we also computed and averaged the corresponding $\tilde{\mathcal{D}}$ values, which are shown as black stars on Fig.~\ref{fig:paretophi}. They lie exactly on the respective Pareto fronts, thus demonstrating that the attainable distance values are determined by the limitation on the control time. 

\begin{table}[htbp]
\caption{\label{t:nophase} Optimization results for 500 D-MORPH runs performed to minimize the phase-independent distance $\mathcal{G}$ of Eq.~(\ref{ind}). The target gates are $W_{\mathrm{QFT},2}$ and $W_{\mathrm{SWAP}}$ with control times $T = 9$ and $T = 10$, respectively. The percentage of searches that approached each target transformation $W(\phi_m) = e^{i \phi_m} W$ is recorded, along with the associated mean value of $\mathcal{G}$. The difference of the reported percentages from $25\%$ ($1/N$ probability) is mostly due to the fact that initial control fields are not completely random, as they are generated using the parameterized form (\ref{field}).}
\begin{ruledtabular}
\begin{tabular}{crcrc}
Gate & $T$ & $\phi_m$ & $\%$ of searches & $\overline{\mathcal{G}}$ \\
\hline
QFT & 9 & 0 & 28.5 & $9.87 \times 10^{-9}$ \\
 &  & $\pi/2$ & 8.6 & $4.77 \times 10^{-3}$ \\
 &  & $\pi$ & 15.6 & $9.91 \times 10^{-9}$ \\
 &  & $3\pi/2$ & 47.3 & $4.71 \times 10^{-3}$ \\
\hline
SWAP & 10 & 0 & 43.3 & $9.95 \times 10^{-9}$ \\
 &  & $\pi/2$ & 26.4 & $2.14 \times 10^{-2}$ \\
 &  & $\pi$ & 22.9 & $9.91 \times 10^{-9}$ \\
 &  & $3\pi/2$ & 7.4 & $2.13 \times 10^{-2}$ \\
\end{tabular}
\end{ruledtabular}
\end{table}

\begin{figure}[t]
\includegraphics[width=7.7cm]{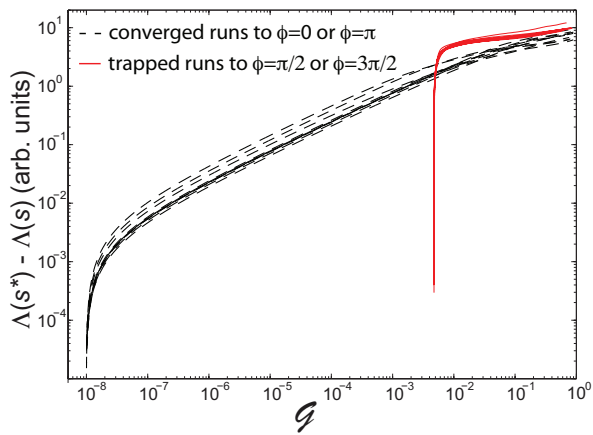}
\caption{(Color online) Path length difference $\Lambda(s^{\star})-\Lambda(s)$ along optimization trajectories seeking $W_\text{QFT,2}$ with $T=9$, as a function of the phase-independent distance ${\mathcal{G}}$. Black (dashed line) trajectories: searches that attained $\mathcal{G} \leq 10^{-8}$ as they converged to $W(\phi_m)$ with $\phi_m =\{0,\pi\}$. Red (solid line) trajectories: searches that were ``trapped'' at $\mathcal{G} \approx 0.0047$ as they approached $W(\phi_m)$ with $\phi_m =\{\pi/2,3\pi/2\}$. Trajectories for the converged searches behave similarly to those with $T>T^{\ast}$ in Fig.~\ref{fig:pathf}. On the other hand, as the red trajectories approach the ``trap'' at $\mathcal{G} \approx 0.0047$, they show a very large change in the field structure ($\Lambda$ changes by $\sim 4$ orders of magnitude) with negligible improvement in fidelity.}
\label{fig:pathglobal}
\end{figure}

Further insight into the difference between searches that approach transformations corresponding to $\phi_m =\{0,\pi\}$ and $\phi_m =\{\pi/2,3\pi/2\}$ can be gained by examining the path length difference $\Lambda(s^{\star})-\Lambda(s)$. From 500 D-MORPH runs listed in Table~\ref{t:nophase} for the gate $W_\text{QFT,2}$ and $T=9$, we selected 20 typical searches: five for each value of $\phi_m$. In Fig.~\ref{fig:pathglobal}, $\Lambda(s^{\star})-\Lambda(s)$ is plotted as a function of $\mathcal{G}$ for these 20 searches. The optimization trajectories corresponding to $\phi_m =\{0,\pi\}$ (shown as black dashed lines) successfully converged ($\mathcal{G} \leq 10^{-8}$) to the respective transformations and exhibit a behavior similar to that of the trajectories with $T>T^{\ast}$ in Fig.~\ref{fig:pathf}. Specifically, $\Lambda(s^{\star})-\Lambda(s)$ changes with $\mathcal{G}$ according to a power law (which appears as a linear change on the log--log plot) over the range $10^{-7} \lesssim \mathcal{G} \lesssim 10^{-3}$. In contrast, the optimization trajectories corresponding to $\phi_m =\{\pi/2,3\pi/2\}$ (shown as red solid lines) halted at a suboptimal objective value $\mathcal{G} \approx 0.0047$, near which the ``trapping'' of these searches is manifested by a very large change in $\Lambda(s)$ (over $\sim 4$ orders of magnitude) that produces negligible improvement in fidelity. These trajectories also exhibit a significant ``flattening'' at $\mathcal{G} \sim 0.1$. The qualitative difference between the trajectories corresponding to the converged searches and those destined to be ``trapped'' due to unreachability of their target transformations can be utilized to identify the latter ones at an early stage of the D-MORPH optimization. We did not observe any search whose trajectory would indicate that it was attracted to an unreachable transformation (corresponding to $\phi_m = \pi/2$ or $3\pi/2$) \emph{en route} to finally converging to a reachable one (corresponding to $\phi_m = 0$ or $\pi$).

We also verified that for sufficiently large $T$ values (i.e., $T \geq 9.86$ for the QFT gate and $T \geq 11.84$ for the SWAP gate), \emph{all} searches attained $\mathcal{G} \leq 10^{-8}$. These results demonstrate that the critical times corresponding to all values of the target gate's global phase have to be accounted for in order to ensure that all relevant target transformations are reachable at the selected control time. From a practical point of view, due to the extremely high search effort in the vicinity of the critical time, it is preferable to use a $T$ value that safely exceeds all $T^{\ast} (\phi_m)$.

\subsection{Dependence of critical time on inter-qubit coupling strength}
\label{sec:2q3}

\begin{figure}[htbp]
\includegraphics[width=7.7cm]{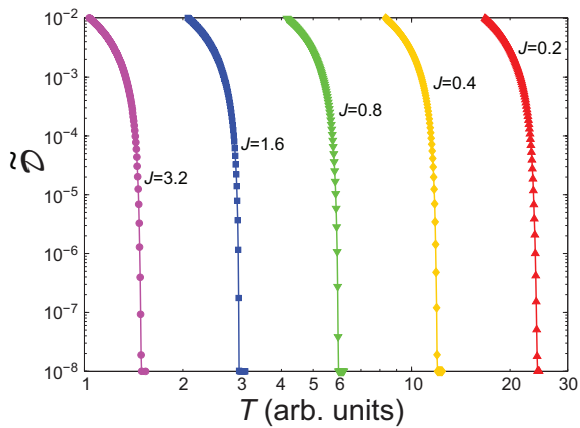}
 \caption{(Color online) Distance-time Pareto fronts for selected values of the inter-qubit coupling strength $J$. The target gate is $W_{\mathrm{QFT},2}$. Each Pareto front is shown by a different symbol shape and color, with the $J$ value indicated to the right of the corresponding curve. The equal spacing between the fronts on the logarithmic-scale abscissa indicates a power law dependence of $T^{\ast}$ on $J$.}
\label{fig:pareto2}
\end{figure}

For two- and multi-qubit gates, the implementation speed is limited, in most practical situations, by strengths of inter-qubit interactions. To explore the relationship between the critical time $T^{\ast}$ and the coupling strength $J \equiv J^{(1,2)}$ for a given target gate $W$, the PFT procedure was performed with various values of $J$ for an otherwise fixed two-qubit Hamiltonian. The resulting distance-time Pareto fronts for systems with $0.2 \leq J \leq 3.2$ and $W = W_{\mathrm{QFT,2}}$ are shown in Fig.~\ref{fig:pareto2}. As $J$ varies, the front's shape remains approximately the same (on the log--log plot), but the entire curve is shifted on the $T$ axis. In Fig.~\ref{fig:pareto2}, the coupling constant is doubled for each successive Pareto front from $J = 0.2$ (the rightmost curve) to $J = 3.2$ (the leftmost curve), and the equal spacing between the fronts on the logarithmic-scale abscissa suggests that $T^{\ast}$ has a power law dependence on $J$ with a negative exponent. 

\begin{figure}[htbp]
\includegraphics[width=7.7cm]{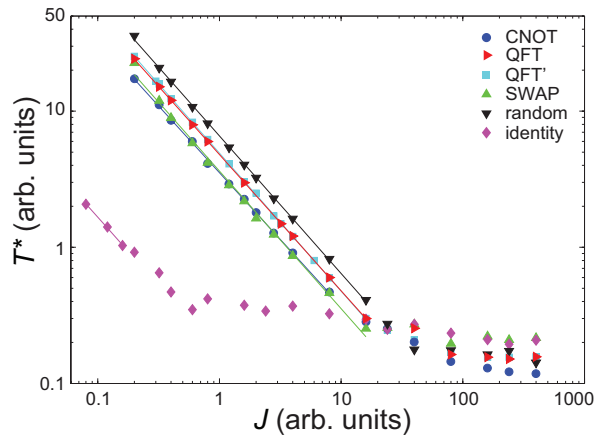}
 \caption{(Color online) Critical times $T^{\ast}$ identified using the PFT method are plotted versus the inter-qubit coupling strength $J$. Points corresponding to different two-qubit target gates are denoted by distinct symbol shapes and colors. Regression lines obtained from the least squares fit have slopes in the range from $-0.980$ to $-1.015$ and are shown over the ranges of $J$ values for which the power-law relation holds. The scaling $T^{\ast} \propto 1/J$ breaks down when $J \gtrsim 0.2$ for the identity transformation, $J \gtrsim 10$ for the CNOT gate, and $J \gtrsim 20$ for the other gates.}
\label{fig:tvsj}
\end{figure}

Figure~\ref{fig:tvsj} confirms that $T^{\ast}$ scales as $1/J$ for a significant range of coupling strength values and various target gates. Specifically, we considered QFT, QFT$'$, CNOT, SWAP, the identity transformation $\openone_4$, and a random transformation $W_{\mathrm{rnd}} \in \mathrm{SU}(N)$ produced by $W_{\mathrm{rnd}} = e^{i A}$, where $A$ is a random, traceless Hermitian matrix. For each of these target gates, the power-law exponent was evaluated as the slope of the corresponding regression line on the log--log plot of Fig.~\ref{fig:tvsj}. These slopes range from $-0.980$ to $-1.015$, in excellent agreement with the scaling $T^{\ast} \propto 1/J$. As the regression lines for various choices of $W$ have different intercepts on the ordinate, the prefactor in the relationship $T = c / J$ is a function of the gate: $c = c(W)$. The value of the prefactor $c$ for the identity transformation is much smaller than for any other gate, while the largest value of $c$ (among the gates considered here) is found for the random transformation.

In previous studies of TOC \cite{Khaneja2001, Khaneja2002, KhanejaReiss2005, KhanejaHeitmann2007, SchulteSporl2005, Carlini2011, Carlini2007}, the scaling $T^{\ast} \propto 1/J$ has been assumed to hold for various two- and multi-qubit gates. However, an explicit numerical validation of this property has not been previously demonstrated over large ranges of $J$ values. Figure~\ref{fig:tvsj} shows that this scaling breaks down when $J \gtrsim 0.01 \omega_1$ for the identity transformation, $J \gtrsim 0.5 \omega_1$ for the CNOT gate, and $J \gtrsim \omega_1$ for the other gates (recall that $\omega_1 = 20$ is the smaller transition frequency of the two qubits). The breakdown of the $1/J$ scaling for non-identity gates happens when inter-qubit and intra-qubit dynamics can occur on roughly the same time scale, i.e., when the rate of the quantum information exchange between qubits stops being the ``bottleneck'' for the speed of the gate implementation.

\begin{figure}[t]
\includegraphics[width=7.7cm]{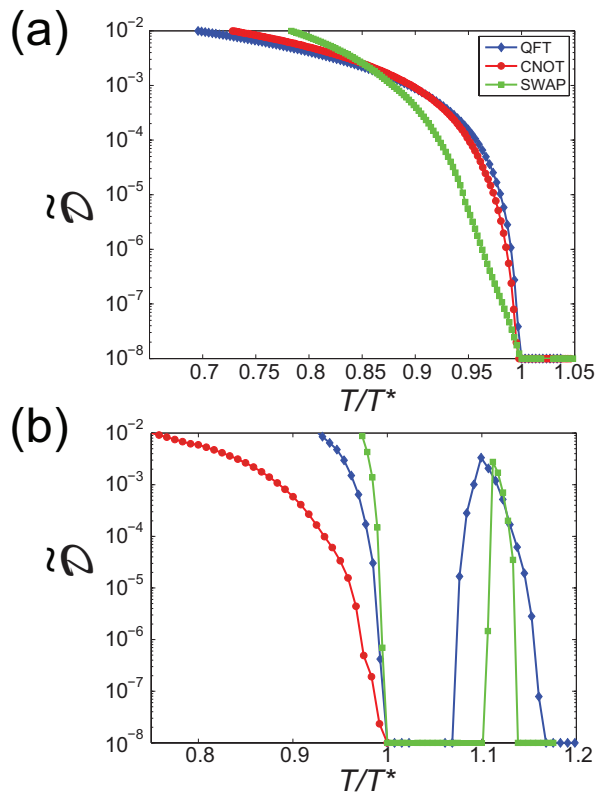}
\caption{(Color online) Distance-time Pareto fronts for the target gates $W_{\mathrm{CNOT}}$, $W_{\mathrm{QFT},2}$, and $W_{\mathrm{SWAP}}$ (denoted by red circles, blue diamonds, and green squares, respectively). The normalized distance $\tilde{\mathcal{D}}$ is plotted versus the reduced time $T/T^{\ast}$. (a) The fronts for $J = 0.8$, where $\tilde{\mathcal{D}} \leq 10^{-8}$ is attainable for all values of $T \geq T^{\ast}$. (b) The fronts for $J = 400$, where the QFT and SWAP transformations are unreachable in a range of time values for $T > T^{\ast}$.}
\label{fig:bigj}
\end{figure}

Distance-time Pareto fronts for strong inter-qubit couplings ($J \gg \omega_1$) exhibit a distinct structure as compared to those for smaller $J$ values. In Fig.~\ref{fig:bigj}, we compare the Pareto fronts for (a) weak coupling: $J = 0.8 = 0.04 \omega_1$, and (b) strong coupling: $J = 400 = 20 \omega_1$, obtained for three target gates (CNOT, QFT, and SWAP). To facilitate the comparison, the distance $\tilde{\mathcal{D}}$ is plotted versus the reduced time $T/T^{\ast}$. When coupling is weak, the  fronts for different target gates have similar, but distinct shapes. When coupling is strong, the fronts for the QFT and SWAP gates exhibit non-monotonic behavior, as the target transformation is unreachable in a range of control times larger than $T^{\ast}$. Such behavior was also observed in other multiobjective optimization studies \cite{Miettinen1998, MostaghimTeich2004}.

According to the data presented in Fig.~\ref{fig:tvsj}, in the strong coupling limit ($J \gg \omega_1$), the critical times $T^{\ast}$ saturate at values on the order of $\pi/\omega_1 \approx 0.157$. A possible explanation for this effect is that, in the considered model systems [c.f., Eqs.~(\ref{ho}) and (\ref{dipole})], one-qubit rotations around the $z$ axis (or, in fact, any arbitrary axis) cannot be accomplished in arbitrarily short time because the control fields are polarized only in the $x$ direction. The obtained values of $T^{\ast}$ suggest that a target transformation would become unreachable when the control time is too short to perform a phase flip (a $\pi$ rotation around the $z$ axis) of the qubit with the smaller transition frequency. A similar situation where the critical time is set by the minimum transition frequency was also encountered in state control \cite{CanevaCalarco2011}. That work \cite{CanevaCalarco2011} also reported an analytical form of the Pareto front, for a variety of controlled quantum systems. The distinct Pareto front shapes shown on Fig.~\ref{fig:bigj} indicate that such a universal analytical form likely does not exist for control of unitary transformations.

\section{Critical times for control of three- and four-qubit gates}
\label{sec:multiq}

In this section, we explore TOC of unitary transformations in three- and four-qubit systems, with the primary goal of identifying critical time values as the number of qubits increases. In particular, we investigate whether the scaling $T^{\ast} \propto 1/J$ that was demonstrated for a range of $J$ values in the two-qubit case also holds for multi-qubit gates. In all simulations reported in this section, we consider the target gate $W_{\mathrm{QFT}',n}$ [Eq.~(\ref{qft})]. In most simulations, the inter-qubit coupling strengths in Eq.~(\ref{ho}) are selected equal, i.e., $J^{(k,j)} = J$, $\forall k,j$. However, several cases of unequal couplings are considered as well for three-qubit systems, with the focus on two physically realistic scenarios: (i) $J^{(k,j)}$ values are similar but not equal (e.g., spins on a lattice or in a polyatomic molecule) and (ii) $J^{(1,2)} = J^{(2,3)} \gg J^{(1,3)}$ (e.g., a linear spin chain). The choices of unequal $J^{(k,j)}$ values along with the average coupling strength $\overline{J} = [J^{(1,2)} + J^{(1,3)} + J^{(2,3)}]/3$ are displayed in Table~\ref{t:js}.

\begin{table}[htbp]
\caption{\label{t:js} Values of inter-qubit coupling strengths $J^{(k,j)}$ employed in simulations with three-qubit systems. The average coupling strength $\overline{J}$ is also shown. The corresponding $T^{\ast}$ values are plotted versus $\overline{J}$ in Fig.~\ref{fig:tvsjq}.}
\begin{ruledtabular}
\begin{tabular}{rrrr}
  $J^{(1,2)}$ & $J^{(1,3)}$ & $J^{(2,3)}$ & $\overline{J}$ \\
  \hline
  2.0 & 1.2 & 1.6 & 1.600 \\
  1.2 & 0.4 & 0.8 & 0.800 \\
  2.0 & 0.4 & 2.0 & 1.467 \\
  2.0 & 0.0 & 2.0 & 1.333 \\
  4.0 & 0.4 & 4.0 & 2.800 \\
  4.0 & 0.0 & 4.0 & 2.667 \\
  8.0 & 0.0 & 8.0 & 5.333 \\
  20.0 & 0.0 & 20.0 & 13.333 \\
  40.0 & 0.0 & 40.0 & 26.667 \\
  80.0 & 0.0 & 80.0 & 53.333 \\
\end{tabular}
\end{ruledtabular}
\end{table}

\begin{figure}[htbp]
\includegraphics[width=7.7cm]{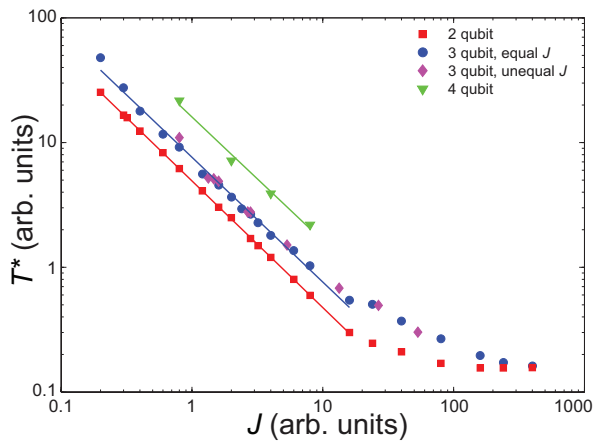}
 \caption{Critical times $T^{\ast}$ plotted versus the inter-qubit coupling strength $J$, for $n$-qubits systems ($n = 2,3,4$). Different system sizes are denoted by distinct symbol shapes and colors. The target gate is $W_{\mathrm{QFT}',n}$. Regression lines obtained from the least squares fit are shown for systems with equal couplings, and their slopes range from $-0.99$ to $-1.015$. The scaling $T^{\ast} \propto 1/J$ breaks down when $J \gtrsim 20$. For three-qubit systems with unequal couplings, $T^{\ast}$ values are plotted versus the average coupling strength $\overline{J}$ (magenta diamonds), and these points fit very well with those obtained for equal couplings (blue circles).}
\label{fig:tvsjq}
\end{figure}

Critical times $T^{\ast}$ obtained from PFT trajectories are plotted in Fig.~\ref{fig:tvsjq} versus the  coupling strength $J$ (for equal couplings) or versus the average coupling strength $\overline{J}$ (for unequal couplings). For two-, three-, and four-qubit systems, the dependence of $\log T^{\ast}$ on $\log J$, in the range of coupling strengths $J < \omega_1$, is evaluated using the least squares fit. The slopes of the resulting regression lines on the log--log plot of Fig.~\ref{fig:tvsjq} are $-1.015 \pm 0.002$ for $n = 2$, $-1.00 \pm 0.02$ for $n = 3$, and $-0.99 \pm 0.07$ for $n = 4$. Thus, in the range of validity, these numerical data agree with the scaling $T^{\ast} \propto 1/J$ very well for each $n$. Also, the $T^{\ast}$ values for three-qubit systems with unequal couplings follow (for $\overline{J} < \omega_1$) the same scaling with respect to $\overline{J}$ (i.e., $T^{\ast} \propto 1/\overline{J}$), and, overall, these points fit very well with those obtained for equal couplings. Thus, critical times for systems with arbitrary $J^{(k,j)}$ values may be reasonably well estimated from equal-coupling results with $J = \overline{J}$. 

The data obtained for both two- and three-qubit systems demonstrate that the scaling $T^{\ast} \propto 1/J$ breaks down for $J \gtrsim \omega_1 = 20$. As $J$ increases beyond $\omega_1$, the $T^{\ast}$ values for two- and three-qubit systems gradually converge and trend towards $\pi/\omega_1 \approx 0.157$. As noted in Sec.~\ref{sec:2q3} above, this value coincides with the minimum time required to perform a phase flip of the qubit with transition frequency $\omega_1$. At $J = 400$ (the strongest coupling that we considered), this critical time value is already reached for $n = 2$ ($T^{\ast} \approx 0.157$) and is very closely approached for $n = 3$ ($T^{\ast} \approx 0.162$). The saturation of $T^{\ast}$ at nearly the same value for both two- and three-qubit systems further suggests that, in the strong coupling limit, the critical time is determined by the minimum time required to accomplish one-qubit rotations.

When the system size increases, the difference between nearby allowed values of the global phase decreases as $|\phi_m - \phi_{m \pm 1}| = 2\pi/N$, and the normalized distance between the corresponding transformations decreases as $\tilde{\mathcal{D}}[W(\phi_m), W(\phi_{m \pm 1})] = \sin^2 (\pi/N)$. Therefore, when the phase-dependent distance $\tilde{\mathcal{D}}$ is employed as the control objective for three- and four-qubit systems, a small fraction of randomly initialized searches may converge to transformations that differ from the target one by a global phase. Indeed, we encountered such a situation in our simulations, when a small number of D-MORPH searches aimed at reaching the target gate $W_{\mathrm{QFT}',3}$ or $W_{\mathrm{QFT}',4}$ converged instead to gates $e^{\pm i \pi/4} W_{\mathrm{QFT}',3}$ or $e^{\pm i \pi/8} W_{\mathrm{QFT}',4}$, respectively. As discussed in Sec.~\ref{sec:2q2}, when such convergences to gates with nearby values of $\phi$ happen, a sufficiently large control time $T \geq \max\{ T^{\ast}(\phi_m) \}$ is required to ensure that all transformations $W(\phi_m)$ are reachable. If this condition on $T$ is not satisfied, at least some local searches will be unable to attain a desired objective value due to the character of the distance-time Pareto front below the critical time. Such a situation where a small fraction of optimization runs employing the BFGS method were seemingly ``trapped'' was reported in Ref.~\cite{FouquieresSchirmer:traps} for a three-qubit spin-chain system and the QFT target gate. This effect is fully explained by the fact that the optimizations in \cite{FouquieresSchirmer:traps} used the $T$ value such that $T^{\ast}(\phi_m) < T < T^{\ast}(\phi_{m'})$ \footnote{The $T^{\ast}(\phi_m)$ values for this three-qubit system and the QFT target gate have been identified in Ref.~\cite{SchulteSporl2005}}, thus making transformations corresponding to some phase values unreachable. We reproduced the results of Ref.~\cite{FouquieresSchirmer:traps} and verified that this spurious ``trapping'' is completely eliminated when the control time $T$ is made sufficiently large.

\section{Conclusions}
\label{sec:conclusion}

This work examined TOC of quantum unitary transformations through the exploration of the Pareto front that quantifies the trade-off between the goals of minimizing the distance to the target gate and the control time. The PFT algorithm was introduced to (1) identify the critical time $T^{\ast}$ below which the target transformation is not reachable and (2) move along the Pareto front to find families of optimal control fields that minimize the distance (or, equivalently, maximize the fidelity) at various values of $T$. Our results suggest that a distinct Pareto front exists for each selection of the control system and the target gate. A practically relevant feature observed for many gates is the strong dependence of the critical time on the global phase of the target transformation. We also examined the dependence of $T^{\ast}$ on the inter-qubit coupling strength $J$ and confirmed the universal scaling $T^{\ast}\propto 1/J$ (or, $T^{\ast}\propto 1/\overline{J}$ for unequal couplings), consistent with expectations in the literature \cite{Khaneja2001, Khaneja2002, KhanejaReiss2005, KhanejaHeitmann2007, SchulteSporl2005, Carlini2011, Carlini2007}. However, we found that, while this scaling holds for a wide range of $J$ values, it breaks down when $J \gtrsim \omega_1$, where $\omega_1$ is the smallest of the qubit transition frequencies in the system. Thus the critical time cannot be made arbitrarily small by increasing the coupling strength, and the ultimate limit on the $T^{\ast}$ value is set by the smallest transition frequency $\omega_1$ which, for the considered model systems, is the minimum speed of one-qubit rotations.

One of the goals of this work was to relate the obtained TOC results to properties of the optimal quantum control landscape. In particular, we observed that, for a given quantum system and a target transformation, different randomly initialized PFT trajectories produce essentially the same Pareto front. This result suggests that (1) as long as $T \geq T^{\ast}$, the favorable landscape topology characteristic for unconstrained control fields remains intact, and (2) for $T < T^{\ast}$, the limitation on the control time significantly affects the value of the landscape's global optimum (so that unit fidelity/zero distance is no longer attainable), but does not lead to landscape fragmentation and formation of local traps (provided that we distinguish between landscapes for target transformations that differ by the global phase). We also found that the optimization search effort rises superexponentially as $T$ decreases and approaches $T^{\ast}$, with corresponding changes observed in metrics quantifying the local structure of the control landscape. In particular, ``flattening" of the landscape near the optimum correlates remarkably well with the search effort growth as the control time decreases.

The structure of the distance-time Pareto fronts identified in this work has important implications for experimental implementation of high-fidelity quantum gates in short times. Taking into account the very steep slope of the Pareto front at $T \lesssim T^{\ast}$ and realistic experimental uncertainties in the gate control time, operating at a $T$ value that safely exceeds the critical times corresponding to all allowed values of the target gate's global phase is desirable to avoid a significant loss of fidelity. Furthermore, values of $T$ very close to $T^{\ast}$ should be avoided in order to keep the optimization search effort at a reasonable level. 

In summary, this work extensively investigated the effects of limiting the time as a resource for optimal control of quantum unitary transformations. While our study employed a particular physical model, the above results are expected to be qualitatively applicable to other coupled-spin systems. In addition to the goal of generating target unitary transformations, the methods presented here can be applied to other objectives in quantum control, including state preparation and optimization of observable expectation values.

\acknowledgments

This work was supported by the Laboratory Directed Research and Development program at Sandia National Laboratories. Sandia National Laboratories is a multi-program laboratory managed and operated by Sandia Corporation, a wholly owned subsidiary of Lockheed Martin Corporation, for the U.S. Department of Energy's National Nuclear Security Administration under contract DE-AC04-94AL85000. H.R. also acknowledges support from IARPA and the ARO.

\bibliographystyle{apsrev4-1}

\end{document}